\documentclass[a4paper,usenatbib,twocolumn]{mn2e}
\usepackage{newtxtext,newtxmath}
\usepackage{lipsum}
\usepackage[T1]{fontenc}
\usepackage{ae,aecompl}
\usepackage{hyperref}
\usepackage{graphicx}	
\usepackage{amsmath}	
\usepackage{amssymb}	
\usepackage{bm}	
\usepackage{caption}
\usepackage{multirow}
\usepackage{multicol}
\usepackage{cuted}
\usepackage{mathtools}
\usepackage{longtable}

\def \apjs {{\rm {ApJS}}}

\def \aap {{\rm {A\&A}}}

\def \apjl{\rm {ApJL}}

\title[Agatha]{Agatha: disentangling periodic signals from correlated noise in a periodogram framework}
\author[F. Feng et al.]
{F. Feng$^{1}$\thanks{E-mail: f.feng@herts.ac.uk or fengfabo@gmail.com}, M. Tuomi$^{1}$, H. R. A.
  Jones$^{1}$\\
$^{1}$Centre for Astrophysics Research, School of Physics, Astronomy and Mathematics, University of Hertfordshire, College Lane, Hatfield AL10 9AB, UK}
\date{\today}

\begin{document}
\maketitle

\begin{abstract}
Periodograms are used as a key significance assessment and visualisation tool to display the significant periodicities in unevenly sampled time series. We introduce a framework of periodograms, called ``Agatha'', to disentangle periodic signals from correlated noise and to solve the 2-dimensional model selection problem: signal dimension and noise model dimension. These periodograms are calculated by applying likelihood maximization and marginalization and combined in a self-consistent way. We compare Agatha with other periodograms for the detection of Keplerian signals in synthetic radial velocity data produced for the Radial Velocity Challenge as well as in radial velocity datasets of several Sun-like stars. In our tests we find Agatha is able to recover signals to the adopted detection limit of the radial velocity challenge. Applied to real radial velocity, we use Agatha to confirm previous analysis of CoRoT-7 and to find two new planet candidates with minimum masses of 15.1\,$M_\oplus$  and 7.08\,$M_\oplus$ orbiting HD177565 and HD41248, with periods of 44.5\,d and 13.4\,d, respectively. We find that Agatha outperforms other periodograms in terms of removing correlated noise and assessing the significances of signals with more robust metrics. Moreover, it can be used to select the optimal noise model and to test the consistency of signals in time. Agatha is intended to be flexible enough to be applied to time series analyses in other astronomical and scientific disciplines. Agatha is available at \url{http://www.agatha.herts.ac.uk}.
\end{abstract}
\section{Introduction}     \label{sec:introduction}
Time-series analyses based on periodograms have been created and developed over decades to satisfy different requirements for the detection of periodic phenomena. To analyze unevenly sampled time-series in the frequency domain, \cite{lomb76} and \cite{scargle82} independently developed the so-called Lomb-Scargle (LS)  periodogram based on a least-squares fit of sinusoids to data. Variations of the LS periodogram have been developed to account for measurement errors \citep{gilliland87, irwin89} or frequency-dependent mean \citep{cumming99, zechmeister09}, nonsinusoidal functions \citep{bretthorst01, cumming04} or multiple periodic signals \citep{anglada12b, baluev13b}. In addition to these LS-like periodograms, Bayesian periodograms have been developed to assess the significance of signals using marginalized likelihoods when assuming uniform prior densities \citep{bretthorst01, mortier14}. 

These periodograms are frequently used in time series analyses in disciplines such as astronomy, climatology, biology and geology. In particular, they are used by astronomers to e.g. detect planetary candidates in the radial velocity (RV) data, to find periodic variations in photometric time series of quasars, and to study asteroseismology. Most periodograms account for the white noise by weighting the data using measurement errors. However, noise in time-series is typically not white but could be correlated in time or even in other dimensions. In the case of Doppler measurements of stars, the RV noise\footnote{Although noise is equivalent to unknown signals, we consider RV variation induced by stellar activity as noise in the context of detecting exoplanets.} is typically correlated in time and wavelength (\citealt{feng17a}; hereafter F17a). Thus, the white-noise periodograms are insufficient in assessing the significance of a periodic signal in red-noise-dominated time series.

Some periodograms have been created to analyze time series contaminated by correlated noise. For example, \cite{schulz02} developed the ``RedFit'' algorithm to fit a first-order autoregressive (AR1) process to paleoclimatic data. However, this periodogram is biased due to the subtraction of the fitted AR1 component from the data rather than fitting the AR1 and sinusoids simultaneously. This problem, caused by subtraction, is frequently mentioned in the field of exoplanet detections (e.g. \citealt{tuomi12c,foreman-mackey15,anglada15}). Recently \cite{hara16} have developed a compressed sensing technique to model the RV red noise as Gaussian process. This periodogram is also biased due to the subtraction of a global mean and/or a noise component from the data during the pre-procession. Moreover, the Gaussian process it employs could interpret signals as noise without proper penalization \citep{feng16}. To remove pointing-induced systematics in the Kepler's two-wheeled extension (K2) data, \cite{angus16} have developed a new periodogram to account for the linear correlation between the target light curve and selected noise proxies. However, this periodogram ignores the time-correlated noise, and does not select the so-called ``Goldilocks noise model'' by optimizing the number of noise proxies, which could be important for avoiding false negatives and positives (\citealt{feng16} and F17a). 

To analyze time-series as complex as encountered with RV data, we introduce ``Agatha''\footnote{Agatha is named after the famous detective novelist, Agatha Christie, for the reason that detecting signals in noise-polluted data is like solving difficult cases in detective fiction.}, a framework of periodogram analyses based on both frequentist and Bayesian methods. Agatha is intended to offer a number of features: (1) fit the time-correlated noise using the moving average model, (2) used to compare noise models to select the Goldilocks noise model \citep{feng16}, (3) optimization of the frequency-dependent linear trend simultaneously with sinusoids and noise components, (4) wavelength-dependent noise accounted for by fitting a set of linear functions of the ``differential RVs'' introduced by F17a to the data, (5) assessment of the significance of signals using the Bayes factor (BF) estimated by the Bayesian information criterion (BIC), which is probably the Goldilocks estimator of BF for the RV data (Feng et al. 2016a), (6) production of the so-called ``moving periodogram'' \citep{feng17b} to visualize the change of signals with time thus visually testing the consistency of signals.

Although periodograms might not be as robust as Bayesian methods implemented through Markov Chain Monte Carlo in selecting and quantifying signals (e.g. \citealt{ford07, fischer16, feng16}), they are computationally efficient and are good at signal visualization. In combination with Bayesian methods, Agatha would greatly improve the efficiency and robustness of signal detections in unevenly sampled time series such as RVs. The code for Agatha is written in R and is available at GitHub: \url{https://github.com/phillippro/Agatha}. A relevant web app is also developed and is available at \url{http://www.agatha.herts.ac.uk}. 

This article is structured as follows. We analytically present the formulae for likelihood-optimization and marginalization to construct the Bayes Factor Periodogram (BFP) and the Marginalized Likelihood Periodogram (MLP) in section \ref{sec:BFP} and in section \ref{sec:MLP}, respectively.  In section \ref{sec:application}, these periodograms are combined to form Agatha, and are compared with other periodograms for selected example RV data sets. Finally, we discuss and conclude in section \ref{sec:conclusion}. 

\section{Bayes factor periodogram}\label{sec:BFP}

We define an unevenly sampled time series as $\{(t_i, v_i)\}$, where $v_i$ is the RV measured at time $t_i$, and $i\in\{1,2,...,N\}$. The basic model used for finding periodic signals in the time series is 
\begin{equation}
  \hat{r}_i=A\cos(2\pi ft_i-\phi)+B\sin(2\pi ft_i-\phi)+\gamma+\dot\gamma t_i +\sum\limits_{j=1}^{N_I}d_jI_{ij}~, 
\label{eqn:model_basic}
\end{equation}
where $f$ is the signal frequency, $\phi$ is an arbitrary phase offset determined by the time reference point\footnote{Since the sine fit is time-translation invariant, the reference time could be arbitrarily chosen without affecting the power of fit.}, $\gamma$ is the intercept, $\dot\gamma$ is the slope characterizing a trend, and ${\bf d}\equiv \{d_j\}$ characterizes the linear dependence of the time series on $N_I$ noise proxies ${\bm I_j}\equiv\{I_{ij}:i\in{1,...,N}\}$. This model is linear with respect to all other parameters but $f$. 

To account for time-correlated noise in the times series, we introduce the moving average (MA) model which is one of the best noise models for the detection of Keplerian signals according to the RV challenge results \citep{tuomi12,dumusque16b}. The full model is 
\begin{equation}
  \hat{v}_i=\hat{r}_i+\sum\limits_{k=1}^qm_k\exp[-|t_i-t_{i-k}|/\tau](v_{i-k}-\hat{r}_{i-k}) ~, 
\label{eqn:model_full}
\end{equation}
where $m_k$ and $\tau$ are the semi-amplitude and time scale of the correlation between data measured at different times, and $(v_{i-k}-\hat{r}_{i-k})$ is the residual of $\hat{r}_{i-k}$ at $t_{i-k}$. Actually, the MA model is a simplified Gaussian process since it only accounts for the correlation between previous data points and the current point. Considering that the Gaussian process may be too flexible to properly disentangle signals from the noise \citep{feng16}, we use $q$ MA components, i.e. MA(q), to model the red noise. Red noise refers to time-correlated noise which does not include wavelength-dependent noise. The MA(0) model with ${\bf d=0}$ is the white noise model which accounts for jitter (or excess white noise) and includes a linear trend. Thus the full model is the white noise model combined with the MA model and the correlation between RVs and noise proxies. 

We assume that the residuals $\{v_i-\hat{v}_i:~i\in\{1, ..., N\}\}$ follow a Gaussian distribution, the likelihood function for model M and data D is 
\begin{equation}
\mathcal{L}({\bm \theta})\equiv P(D|{\bm \theta},M) = \prod_i\frac{1}{\sqrt{2\pi(\sigma_i^2+\sigma_J^2)}}\exp\left[-\frac{(v_i-\hat{v}_i)^2}{2(\sigma_i^2+\sigma_J^2)}\right]~,
\label{eqn:likelihood}
\end{equation}
where $\bm \theta$ is the model parameters, $\sigma_J$ is a parameter used to model the so-called ``jitter'' in the time series, and $\sigma_i$ is the known measurement error of $v_i$. In reality, we estimate the optimal values of parameters by maximizing the natural logarithm of the likelihood which is 
\begin{equation}
  \ln{\mathcal{L}(\bm \theta)}= -\sum\limits_i\frac{ \ln[2\pi(\sigma_i^2+\sigma_J^2)]}{2}-\sum\limits_i\frac{(v_i-\hat{v}_i)^2}{2(\sigma_i^2+\sigma_J^2)}~.
  \label{eqn:logLike}
\end{equation}

However, it is computationally expensive to directly maximize the logarithmic likelihood as a function of many parameters. To make the parameter optimization more efficient, we first express the logarithmic likelihood as a function of ${\bf m}=\{m_k\}$, $\tau$ and $\sigma_J$ by analytically maximizing the logarithmic likelihood. We then use the R package ``minpack.lm''\footnote{This package is available at \href{https://CRAN.R-project.org/package=minpack.lm}{https://CRAN.R-project.org/package=minpack.lm}.} to maximize the logarithmic likelihood as a function of $\bf m$, $\tau$ and $\sigma_J$. To simplify the calculation, we introduce the following notations,
\begin{eqnarray}
  c_{ik}&=&m_k\exp(-|t_i-t_{i-k}|/\tau)~,\\
  v_i'&=&v_i-\sum\limits_{k=1}^qc_{ik}v_{i-k}~.
\end{eqnarray}
The residual after subtracting $\hat{v}_i$ from $v_i$ is
\begin{eqnarray}
  \epsilon_i&=&v_i'-A\left[\cos(2\pi f t_i-\phi)-\sum\limits_{k=1}^{q}c_{ik}\cos(2\pi f t_{i-k}-\phi)\right]\nonumber\\
            &&-B\left[\sin(2\pi f t_i-\phi)-\sum\limits_{k=1}^qc_{ik}\sin(2\pi f t_{i-k}-\phi)\right]-\gamma(1-\sum\limits_{k=1}^qc_{ik})\nonumber\\
 &&-\dot\gamma(t_i-\sum\limits_{k=1}^qc_{ik}t_{i-k})-\sum_{j=1}^{N_I}d_j(I_{ij}-\sum\limits_{k=1}^qc_{ik}I_{i-k,j})~.
                \label{eqn:residual}
\end{eqnarray}
We further denote
 \begin{equation*}
  \begin{aligned}[c]
    w'_i&=1-\sum\limits_{k=1}^qc_{ik}~,\\
    c'_i&=\cos(2\pi f t_i-\phi)-\sum\limits_{k=1}^qc_{ik}\cos(2\pi f t_{i-k}-\phi)~,\\
    s'_i&=\sin(2\pi f t_i-\phi)-\sum\limits_{k=1}^qc_{ik}\sin(2\pi f t_{i-k}-\phi)~,\\
    t'_i&=t_i-\sum\limits_{k=1}^qc_{ik}t_{i-k}~,\\
    I'_{ij}&=I_{ij}-\sum\limits_{k=1}^qc_{ik}I_{i-k,j}~,\\
    W'&=\sum\limits_i\omega_iw'_i~,\\
    W&=\sum\limits_i1/(\sigma_i^2+\sigma_J^2)~,\\
    \widehat{YY}&=\sum\limits_{i}\omega_iv_i'v'_i~,\\
    \widehat{YC}&=\sum\limits_{i}\omega_iv_i'c'_i~,\\
  \end{aligned}
\end{equation*}
and
\begin{equation*}
  \begin{aligned}[c]
    \widehat{YS}&=\sum\limits_{i}\omega_iv_i's'_i~,\\
    \widehat{YW}&=\sum\limits_{i}\omega_iv_i'w'_i~,\\
    \widehat{YT}&=\sum\limits_{i}\omega_iv_i't'_i~,\\
    \widehat{YI_j}&=\sum\limits_{i}\omega_iv_i'I'_{ij}~,\\
    \widehat{YI_j}&=\sum\limits_{i}\omega_iv_i'I'_{ij}~,\\
    \widehat{CC}&=\sum\limits_{i}\omega_i{c'}_i^2~,\\
    \widehat{CS}&=\sum\limits_{i}\omega_ic'_is'_i~,\\
    \widehat{CW}&=\sum\limits_{i}\omega_ic'_iw'_i~,\\
    \widehat{CT}&=\sum\limits_{i}\omega_ic'_it'_i~,\\
    \widehat{CI_j}&=\sum\limits_{i}\omega_is'_iI'_j~,\\
  \end{aligned}
  \quad\quad
  \begin{aligned}[c]
    \widehat{SS}&=\sum\limits_{i}\omega_i{s'}_i^2~,\\
    \widehat{SW}&=\sum\limits_{i}\omega_is'_iw'_i~,\\
    \widehat{ST}&=\sum\limits_{i}\omega_is'_it'_i~,\\
    \widehat{SI_j}&=\sum\limits_{i}\omega_is'_iI'_j~,\\
    \widehat{WW}&=\sum\limits_{i}\omega_i {w'}_i^2~,\\
    \widehat{WT}&=\sum\limits_{i}\omega_i w'_it'_i~,\\
    \widehat{WI_{j}}&=\sum\limits_{i}\omega_iw'_iI'_{ij'}~,\\
    \widehat{TT}&=\sum\limits_{i}\omega_i{t'}_i^2~,\\
    \widehat{TI_j}&=\sum\limits_{i}\omega_it'_iI'_{ij}~,\\
    \widehat{I_jI_{j'}}&=\sum\limits_{i}\omega_iI'_{ij}I'_{ij'}~,
  \end{aligned}
\end{equation*}
where $\omega_i=1/((\sigma_i^2+\sigma_J^2)W)$ is the normalized weighting function.
 
Since the fit of the periodic model is time-translation invariant \citep{scargle82}, we adopt $\phi=0$ for the optimization of parameters. By maximizing the logarithmic likelihood in Eqn. \ref{eqn:logLike}, we find the following equation, 
\begin{equation}
  \begin{aligned}
& \left(
   \begin{array}{c*6{c}}
      \widehat{CC}&\widehat{CS}&\widehat{CW}&\widehat{CT}&\widehat{CI_1}&...&\widehat{CI_{N_I}}\\
      \widehat{CS}&\widehat{SS}&\widehat{SW}&\widehat{ST}&\widehat{SI_1}&...&\widehat{SI_{N_I}}\\
\widehat{CW}&\widehat{SW}&\widehat{WW}&\widehat{WT}&\widehat{WI_1}&...&\widehat{WI_{N_I}}\\
      \widehat{CT}&\widehat{ST}&\widehat{WT}&\widehat{TT}&\widehat{TI_1}&...&\widehat{TI_{N_I}}\\
      \widehat{CI_1}&\widehat{SI_1}&\widehat{WI_1}&\widehat{TI_1}&\widehat{I_1I_1}&...&\widehat{I_1I_{N_I}}\\
      \vdots&\vdots&\vdots&\vdots&\vdots&\ddots&\vdots\\
      \widehat{CI_{N_I}}&\widehat{SI_{N_I}}&\widehat{WI_{N_I}}&\widehat{TI_{N_I}}&\widehat{I_1I_{N_I}}&...&\widehat{I_{N_I}I_{N_I}}
    \end{array}  
  \right)\\
  &\times
  \left(
\begin{array}{c}
      A\\
B\\
\gamma\\
\dot\gamma\\
d_1\\
\vdots\\
d_{N_I}\\
    \end{array}  
  \right)=
  \left(
    \begin{array}{c}
      \widehat{YC}\\
\widehat{YS}\\
\widehat{YW}\\ 
\widehat{YT}\\
\widehat{YI_1} \\
\vdots\\
\widehat{YI_{N_I}} \\
    \end{array}  
    \right)~.
      \end{aligned}
  \label{eqn:optwhite}
\end{equation}
By solving the above equations, we obtain the model parameters as functions of  ${\bf m}$, $\tau$ and $\sigma_J$. Then the residual expressed in Eqn. \ref{eqn:residual} and the likelihood in Eqn. \ref{eqn:likelihood} are functions of ${\bf m}$, $\tau$ and $\sigma_J$. We use the Levenberg-Marquardt (LM) optimization algorithm \citep{levenberg44,marquardt63} in order to maximize the logarithmic likelihood to obtain the optimized model parameters and the maximum likelihood.

To assess the significance of signals, we follow \cite{feng16} to estimate the Bayes factor (BF) using the BIC which is equivalent to the maximum likelihood ratio of the periodic model and the noise model. Specifically, we calculate the maximum likelihood for the noise model (i.e. $A=B=0$) and for the full model for a given period. Then we calculate the BIC according to
\begin{equation}
  {\rm BIC}=-2\ln{\mathcal L}_{\rm max}+n\ln{N}~, 
   \label{eqn:BIC}
  \end{equation}
  where ${\mathcal L}_{\rm max}$ is the maximum likelihood, $n$ is the number of free parameters and $N$ is the number of data points (see \citealt{kass95} for details). We calculate the logarithmic BF for a given period using
  \begin{equation}
   \ln{{\rm BF}_{10}}=\frac{{\rm BIC}_0-{\rm BIC}_1}{2}~,
   \label{eqn:BF}
 \end{equation}
 where ${\rm BIC}_1$ and ${\rm BIC}_0$ are BICs for the periodic model and noise model, respectively. If $\ln{\rm BF}_{10}$ is larger than 5, the periodic model for a given period is favored over the noise model and the signal at this period is considered significant. This criterion is based on a comparison of many BF estimators by \cite{feng16} and is also recommended by \citealt{kass95}.

 To calculate the BFP, we evenly sample the frequency from a uniform distribution over $[1/\Delta t_{\rm max}, 1/\Delta t_{\rm min}]$ with a step of $1/\Delta t_{\rm max}$. In this work, we set $\Delta t_{\rm min}=1$\,day\footnote{For synthetic data sets, we set $\Delta t_{\rm min}$ slightly larger than 1 to avoid aliases since the real signals are known to us.} and $\Delta t_{\rm max}$ to be the time span of the data, leading to an oversampling of Nyquist frequency typically by a factor of more than 10. The values of these parameters could be changed for different applications. Here we primarily investigate known signals with periods of longer than a day and prefer to avoid the strong aliases around 1 day. We then calculate $\ln({\rm BF})$ for each frequency/period, and construct the BFP from these logarithmic BFs. 
  
\section{Marginalized likelihood periodogram}\label{sec:MLP}
Although the BFP penalizes model complexity by applying BIC-estimated BF, it assumes a Gaussian-like posterior for each parameter and treats each parameter equally as a free parameter of the model. But such assumptions are not always valid, especially when the posterior is multimodal. According to the Bayesian theorem, the posterior distribution of parameters $\bm \theta$ for a given model $M$ is
\begin{equation}
P({\bm \theta}| D, M)=\frac{P(D|{\bm \theta},M)P({\bm \theta}|M)}{P(D|M)},
\end{equation}
where $P(D|M)=\int P(D|{\bm \theta},M)P({\bm \theta}|M) d {\bm \theta}$ is the so-called ``evidence'' or integrated likelihood, $P(D|{\bm \theta},M)$ is the likelihood function, and $P({\bm \theta}|M)$ is the prior probability density. The evidence ratio of two models is the Bayes factor. 

Assuming uniform prior distributions for all parameters, the posterior of frequency $f$ is \begin{equation}
  P(f|D,M)\equiv\int P({\bm \theta'},{\bm \theta_{\rm fix}},f| D,  M) d {\bm \theta'}\propto \int \mathcal{L}({\bm \theta'}, {\bm \theta_{\rm fix}},f) d{\bm \theta'} ~,
\end{equation}
where $\bm \theta'$ are the parameters to be marginalized, namely ${\bm \theta}'=\{A, B, \gamma, \dot\gamma\}$, and ${\bm \theta}_{\rm fix}\equiv \{{\bf d}, {\bf m}, \tau, \sigma_J\}$ are the parameters which are determined by the BFP without including sinusoidal functions in the model. Since the integral of likelihood over ${\bm \theta}_{\rm fix}$ cannot be calculated analytically, we either fix these parameters at their optimized values estimated by the BFP or subtract the BFP-determined noise component (excluding the trend) from the data. We will present the formulae for the former method, and then set ${\bf d=0}$, ${\bf m=0}$ to obtain the formulae for the latter. 

We drop the non-exponential term in the expression of the likelihood (see Eqn. \ref{eqn:likelihood}) because only the relative significance of periodic signals is relevant. Then the posterior becomes
\begin{equation}
  P(f|D,M)\propto \mathcal{E}_{\bm \theta'}\equiv \int \limits_{\bm \theta'} \exp\left[-\frac{1}{2}\sum_i\frac{\epsilon_i^2}{\sigma_i^2+\sigma_J^2}\right]~,
  \label{eqn:integral}
  \end{equation}
where $\epsilon_i=v_i-\hat{v}_i$. Following \cite{mortier14}, we eliminate the term $\widehat{CS}$ by setting the phase to be\footnote{Since ${\bf d}$, ${\bf m}$, $\tau$ are determined by the BFP without using sinusoidal functions, the phase $\phi$ in the MLP could be different to the value used in the calculation of BFP.}
\begin{equation}
\phi=\frac{1}{2}\tan^{-1}(\frac{2\widehat{C'S'}}{\widehat{C'C'}-\widehat{S'S'}})~,
\end{equation}
where 
\begin{eqnarray}
  \widehat{C'C'}&=&\sum\limits_i\omega_i\left[\cos(2\pi f t_i)-\sum\limits_{k=1}^qc_{ik}\cos(2\pi f t_{i-k})\right]^2~,\\
  \widehat{C'S'}&=&\sum\limits_i\omega_i\left[\cos(2\pi f t_i)-\sum\limits_{k=1}^qc_{ik}\cos(2\pi f t_{i-k})\right]\nonumber\\
  &&\times\left[\sin(2\pi f t_i)-\sum\limits_{k=1}^qc_{ik}\sin(2\pi f t_{i-k})\right]~,\\ \widehat{S'S'}&=&\sum\limits_i\omega_i\left[\sin(2\pi f t_i)-\sum\limits_{k=1}^qc_{ik}\sin(2\pi f t_{i-k})\right]^2~.
\end{eqnarray}

This gives us
\begin{eqnarray}
-\frac{1}{2}\sum_i\frac{\epsilon_i^2}{\sigma_i^2}&=&W\{-\frac{1}{2}[\widehat{YY}+A^2 \widehat{CC}+B^2 \widehat{SS}+\gamma^2 \widehat{WW}+\dot\gamma^2 \widehat{TT}] \nonumber\\
  && + A\widehat{YC}+B \widehat{YS}
     +\gamma \widehat{YW}+\dot\gamma \widehat{YT}-A\gamma\widehat{CW}-A\dot\gamma\widehat{CT} \nonumber\\ &&-B\gamma\widehat{SW}-B\dot\gamma\widehat{ST}-\gamma\dot\gamma\widehat{WT}\}~. 
  \label{eqn:logL}
\end{eqnarray}    

Since the above integrand is the sum of second-degree polynomials of free parameters,  we can integrate the integrand with respect to parameter $x$ by expressing it as $ax^2+bx+c$, where $a$, $b$ and $c$ are functions of the model parameters. Following \cite{mortier14} and by repeatedly using formula $\int_{-\infty}^{\infty} \exp(ax^2+bx)=\sqrt{\frac{\pi}{|a|}}\exp(-b^2/4a)$, the integral in Eqn. \ref{eqn:integral} becomes
\begin{equation}
  \centering 
\begin{aligned}
    \mathcal{E}_{\bm \theta'}(f) &=\frac{(2\pi)^2}{W^2\sqrt{|V|}}\exp\left[ \frac{W}{2\widehat{CC}\widehat{SS}U}(X+\frac{G^2}{V})\right]~,
    \end{aligned}
  \label{eqn:evi_full}
\end{equation}
 where
\begin{eqnarray*}
G&=&\widehat{CC}\widehat{SS}\widehat{YT}U-\widehat{YC}\widehat{CT}\widehat{SS}U-\widehat{YS}\widehat{ST}\widehat{CC}U+\widehat{CC}\widehat{SS}QR\\
V&=&\widehat{CC}\widehat{SS}\widehat{TT}U- \widehat{SS}\widehat{CT}^2U-\widehat{CC}\widehat{ST}^2U-\widehat{CC}\widehat{SS}R^2\\
X&=&\widehat{SS}\widehat{YC}^2U+\widehat{CC}\widehat{YS}^2U-\widehat{YY}\widehat{CC}\widehat{SS}U+\widehat{CC}\widehat{SS}Q^2\\
U&=& \widehat{WW} -\widehat{CW}^2/\widehat{CC}-\widehat{SW}^2/\widehat{SS}\\
Q&=&\widehat{YW}-\widehat{CW}\widehat{YC}/\widehat{CC}-\widehat{SW}\widehat{YS}/\widehat{SS}\\
R&=&\widehat{CW}\widehat{CT}/\widehat{CC}+\widehat{SW}\widehat{ST}/\widehat{SS}-\widehat{WT}~.
\end{eqnarray*}  

The marginalized posterior/likelihood ratio of the models with frequency $f_1$ and $f_2$ is
\begin{equation}
  \frac{P(f_1|D,M)}{P(f_2|D,M)}=\frac{\mathcal{E}_{\bm\theta'}(f_1)}{\mathcal{E}_{\bm\theta'}(f_2)}~. 
\end{equation}
Following \cite{mortier14}, we scale the marginalized likelihoods (MLs) to their maximum value to define the relative  ML, namely $\rm ML/ML_{max}$. Since the likelihood is only marginalized over a limited number of parameters, the relative ML is not Bayes factor, and is thus not appropriate for assessing signal significance.

As mentioned before, there are two ways to construct an MLP. One method is to optimize the noise model and fix ${\bf m}$, $\tau$, $\bf d$ and $\sigma_J$ at the optimal values, and calculate the relative ML. This approach is called the ``parameter-fixed'' method. The other method is to optimize the noise model and subtract the optimal model prediction from the data, fix $\sigma_J$ at the optimal value, set ${\bf m=d=0}$, and calculate the relative probability. This approach is called the ``noise-subtracted'' method. However, both methods are biased because the parameters of the correlated noise component are determined for the null hypothesis and are not marginalized simultaneously with other parameters, probably leading to an underestimation of signal significance. This bias will be discussed in section \ref{sec:dRV}. We calculate the MLP using the noise-subtracted method keeping the parameter-fixed method as an option.

\section{Application of Agatha}\label{sec:application}
We combine the BFP and MLP to form Agatha which is able to compare noise models, fit the correlated noise and test the consistency of signals in time. Although the BFP and MLP can be used independently, we suggest to use them in combination with Bayesian methods to analyze irregular time series in the following way. 
\begin{itemize}
\item Select the Goldilocks noise model through optimizing the model parameters by using Eq. \ref{eqn:optwhite} and the LM algorithm, and calculate the logarithmic BF using Eqs. \ref{eqn:BIC} and \ref{eqn:BF} for model comparison (see section \ref{sec:comparison}). 
\item Calculate the BFP to select the signals with the highest logarithmic BF (see sections \ref{sec:index}, \ref{sec:ma} and \ref{sec:dRV} for examples). 
\item Use the selected signal as a guidance for the search of signals by applying posterior sampling implemented by the Markov Chain Monte Carlo (MCMC) method. 
\item Estimate the parameters for the selected signal based on posterior sampling.
\item Calculate the BIC-estimated logarithmic BF of $k$-planet model and $k-1$-planet. If logarithmic BF is larger than 5, subtract the best-fitted Keplerian components from the data and calculate the residual BFP to identify the next potential signal (see sections \ref{sec:ma} and \ref{sec:dRV}).
\item Repeat the above three steps until $\ln{\rm BF}$ is less than 5 and there is no significant signals in the residual BFP. 
\item Test the consistency of all signals using the BFP-based/MLP-based moving periodogram (see section \ref{sec:mp}). 
\end{itemize}
To test the validation of Agatha and quantify the signals identified by Agatha, we use the Bayesian method implemented by posterior sampling through the adaptive Metropolis (AM) algorithm \citep{haario06}. We use uniform prior over the logarithmic scale for time parameters and use uniform priors for other parameters. Following \cite{tuomi12b}, we confirm the existence of a signal if the posterior distribution over the period can be constrained from above and below. We also calculate the BIC-based BFs and adopt a logarithmic BF threshold of 5 to select signals. The reader is referred to \cite{tuomi12} and \cite{feng16} for details of the AM method. In the following subsections, we will specify how Agatha is used for different applications in data analyses of RV data. 

\subsection{Data}\label{sec:data}
We have investigated a number of different radial velocity datasets with Agatha during its development, for example \cite{feng17b}. Here we select example synthetic and real RV data sets in order to present the algorithms and methodology behind the usage of Agatha.

To see the improvement of periodograms by inclusion of activity indices, we compare periodograms for the 492 synthetic data points of the second RV challenge data set due to the strong correlation between RVs and indices and because the injected 75.28 day signal is at the limits of detectability \citep{dumusque16a}. It has also been analyzed using compressed sensing techniques by \cite{hara16}. Five signals corresponding to the five planets in the Kepler-20 system are injected into simulated noise sampled according to the observational calendar of HARPS measurements of $\tau$ Ceti. The periods of these signals are 3.77, 5.79, 10.64, 20.16 and 75.28\,d with semi-amplitudes of 2.75, 0.27, 2.85, 0.34 and 1.35\,m/s. The simulated rotation period is 25.05\,d.

To see the difference between red-noise and white-noise periodograms, we choose the 221 HARPS RVs of HD41248 because this data set has been the subject of some debate in the literature (e.g., \citealt{jenkins13};  \citealt{jenkins14}; \citealt{santos14}). It presents a good sized dataset with a reasonable sampling of observational times. The data for HD41248 is essentially the same as that used by \cite{jenkins14} although we reprocessed using the TERRA algorithm \citep{anglada12} and make use of the wavelength dependent data, which is available in the appendix. 

The dataset for HD177565 is chosen as a fairly typical in terms of number of points and phase coverage and with no signals previously reported. It is chosen to illustrate the necessity of modeling wavelength-dependent noise in the detection of weak signals. We present the 68 HARPS data points of HD177565 in Appendix \ref{sec:HD177565}. The data is reduced using the TERRA algorithm which produces RVs for each individual spectral order.

To model the wavelength-dependent noise, F17a have linearly included the RV differences between spectral orders into the model to be used as noise proxies. Specifically, we evenly divide the spectral orders into groups and average the orders in each group to form the so-called ``aperture data sets''. For $n$-summations of orders, we thus create $n$ aperture data sets denoted by ``$n$AP$i$'', where $i\in \{1, 2, ..., n\}$. Then the difference between aperture data sets are called ``differential RVs'', named by ``$n$AP$i$-$j$'', where $i\in \{2, ..., n\}$ and $j=i-1$.

To remove the instrumental bias of HARPS, we generate aperture data sets from a set of 168 HARPS data sets measured for different targets, remove the outliers and stack them to generate the so-called ``calibration data sets'' (F17a). We derive aperture data sets and differential RVs from these HARPS measurements, and combine them. Specifically, we remove the (differential) RVs which have absolute values larger than 20 m/s or deviate from the mean more than 5$\sigma$ before combining them. For each epoch in each aperture data set for a target, we average the calibration (differential) RVs measured within the same night by weighting them according to their measurement errors. We further remove the outliers which deviate from the mean more than 3$\sigma$. There are also epochs where no RVs of other stars are available, we assign the (differential) RVs measured at nearby epochs to them. We use these calibration data sets as proxies like activity indices to remove instrumental noise. For example, we can use a linear combination of the 1AP1, 3AP2-1 and 3AP3-2 calibration data sets to model the instrumental noise in the 1AP1 data set. We use ``cnAP$i$'' and ``cnAP$i$-$j$'' to denote the nAP$i$ and nAP$i$-$j$ calibration data sets. Since the c3AP2-1 data set is found to be strongly correlated with RVs, it is linearly included into the full model in Eqn. \ref{eqn:model_full}. Hereafter, we use this calibration data in the noise model for real RV data sets.

\subsection{Model comparison}\label{sec:comparison}
Using the BF as a metric, Bayesian inference can be used to compare models on the same footing. However, because the posterior is typically complex and multimodal, it is difficult to calculate the BF analytically. Therefore the BF is usually calculated in a Monte Carlo fashion by posterior sampling. In previous work, we have used the posterior samplings to decide which noise model is the optimal one for modeling RV noise (F17a and \citealt{feng17b}). The posterior distributions for noise-model parameters are typically unimodal according to our analyses. Thus the BIC-estimated BF is probably a good approximation of the true value of BF. The BIC-estimated BF is also the most conservative and efficient BF estimator according to the comparison of different BF estimators for synthetic and real RV data sets \citep{feng16}. Therefore, based on the BIC-estimated BFs, the BFP is an efficient and valid inference tool for noise model comparison.

Following F17a, we compare noise models with different numbers of MA components ($q$) and differential RVs ($N_D\equiv N_{\rm AP}-1$ where $N_{\rm AP}\in\{1, 3, 6, 9, 18, 72\}$)\footnote{If $N_{\rm AP}=1$, the spectral orders are averaged to form one aperture data set. Therefore there is no differential RVs or $N_D=0$. }. For a noise model with a given number of MA components and differential RVs, we adopt different initial values for each parameter, maximize the logarithmic likelihood for each initial value set using the LM algorithm, and select the highest likelihood to be used when calculating the BF according to Eqs. \ref{eqn:BIC} and \ref{eqn:BF}.

To find the global likelihood maxima rather than the local maxima in the likelihood distribution, we select initial values of $\ln{\tau}$ according to a uniform prior with boundaries determined by the minimum difference between observation times and the time span of the time series. For the other parameters, we select the initial values from uniform distributions over intervals determined either by the data or by our prior knowledge. For example, we vary $d_j$ according to a uniform prior distribution over $[-d_{\rm max}, d_{\rm max}]$ with $d_{\rm max}=2(v_{\rm max}-v_{\rm min})/(I_{j,  {\rm max}}-I_{j, {\rm min}})$, where $v_{\rm max}$ and $v_{\rm min}$ are the maximum and minimum of RVs, and $I_{j,  {\rm max}}$ and $I_{j,  {\rm min}}$ are the maximum and minimum of $I_j$, respectively. The reader is referred to \cite{feng16} for more details of prior distributions of parameters, although the numerical solution of the maximum likelihood is not sensitive to prior choices. The number of generated initial values is equal to the rounding of $10+10(N_{\rm AP}/3+2q)$. 

Considering the flexibility of the MA model \citep{feng16}, we use a logarithmic BF threshold of 5 to select $q$. Since the dependence of the data on differential RVs is linear and thus is not as flexible as the MA components, we use a threshold of 2.3 to select $N_{\rm AP}$. We calculate the BFs with the BFP and AM methods for the HARPS data of HD41248 and HD177565. The linear correlation between RVs and BIS, FWHM and S-index are included in the noise models. We report the BFs with respect to the white noise model (i.e. $q=0$ and $N_{\rm AP}=1$) in Table \ref{tab:BFs}. We find that the logarithmic BFs calculated using the BFP and AM typically differ less than 1, confirming the validation of using the BFP as a model comparison tool. According to the BF thresholds we have mentioned, the optimal numbers of MA components and differential RVs are $\{q, N_{\rm AP}\}=\{1,1\}$ and $\{1,3\}$ for HD41248 and HD177565, respectively. We will apply these Goldilocks noise models to the HARPS data in Sections \ref{sec:ma} and \ref{sec:dRV}. In principle, the activity indices of BIS, FWHM and S-index can also be compared in a similar fashion. But to be simple, we combine all of them with differential RVs linearly in the following subsections. We also include c3AP3-2 linearly in the model in section \ref{sec:ma}, \ref{sec:dRV} and \ref{sec:mp}.
\begin{table}
  \centering 
  \caption{The logarithmic BFs of noise models calculated using the BFP and AM methods for the HARPS data set of HD41248 and HD177565. The BFs are calculated with respect to the noise model with $\{q, N_{\rm AP}\}=\{0,1\}$. The BFs of the optimal noise models are shown in boldface. }
\label{tab:BFs}
\begin{tabular}{cc | *{2}{c} | cc | *{2}{c}}
  \hline 
  \multicolumn{4}{c | }{HD41248}&\multicolumn{4}{c}{HD177565}\\\hline 
\multicolumn{2}{c|}{Model}&\multicolumn{2}{c|}{Method}&\multicolumn{2}{c|}{Model}&\multicolumn{2}{c}{Method}\\\hline 
$q$&$ N_{\rm AP}$ & BFP& AM&$q$&$ N_{\rm AP}$ & BFP& AM\\\hline 
0 &1&0&0 &0 &1&0&0\\
0 &3&-0.781&-1.40&0&3&12.6&11.8\\
0&6&-5.4&-6.50&0&6&6.96&6.5\\\hline 
1 &1&{\bf 38.4}&{\bf 40.3}&1&0&13.2&14.9\\
1 &3&37.9&37.2&1&3&{\bf 23.1}&{\bf 22.6}\\
1&6&30.9&29.1&1&6&19.4&18.2\\\hline 
2 &1&40.7&42.6&3&0&11.1&12.8\\
2 &3&39.0&37.5&3&3&21.6&21.5\\
2 &6&32.2&30.4&3&6&17.8&17.1\\\hline 
 \end{tabular}
\end{table}

\subsection{Periodograms for index-dependent noise}\label{sec:index}
Periodogram analysis without accounting for the dependence of RVs on activity indices would be misleading if the dependence was strong. Although the linear dependence can be removed from the data before periodogram analysis, the subtraction is biased due to a lack of simultaneous fitting of noise and signals, as demonstrated visually by \cite{anglada15}. We explain this in detail in section \ref{sec:dRV}. The BFP is able to avoid such a bias by optimizing the parameters of noise and signal for each frequency and can thus better determine the maximum likelihood.

\begin{figure*}
\centering
\includegraphics[scale=0.6]{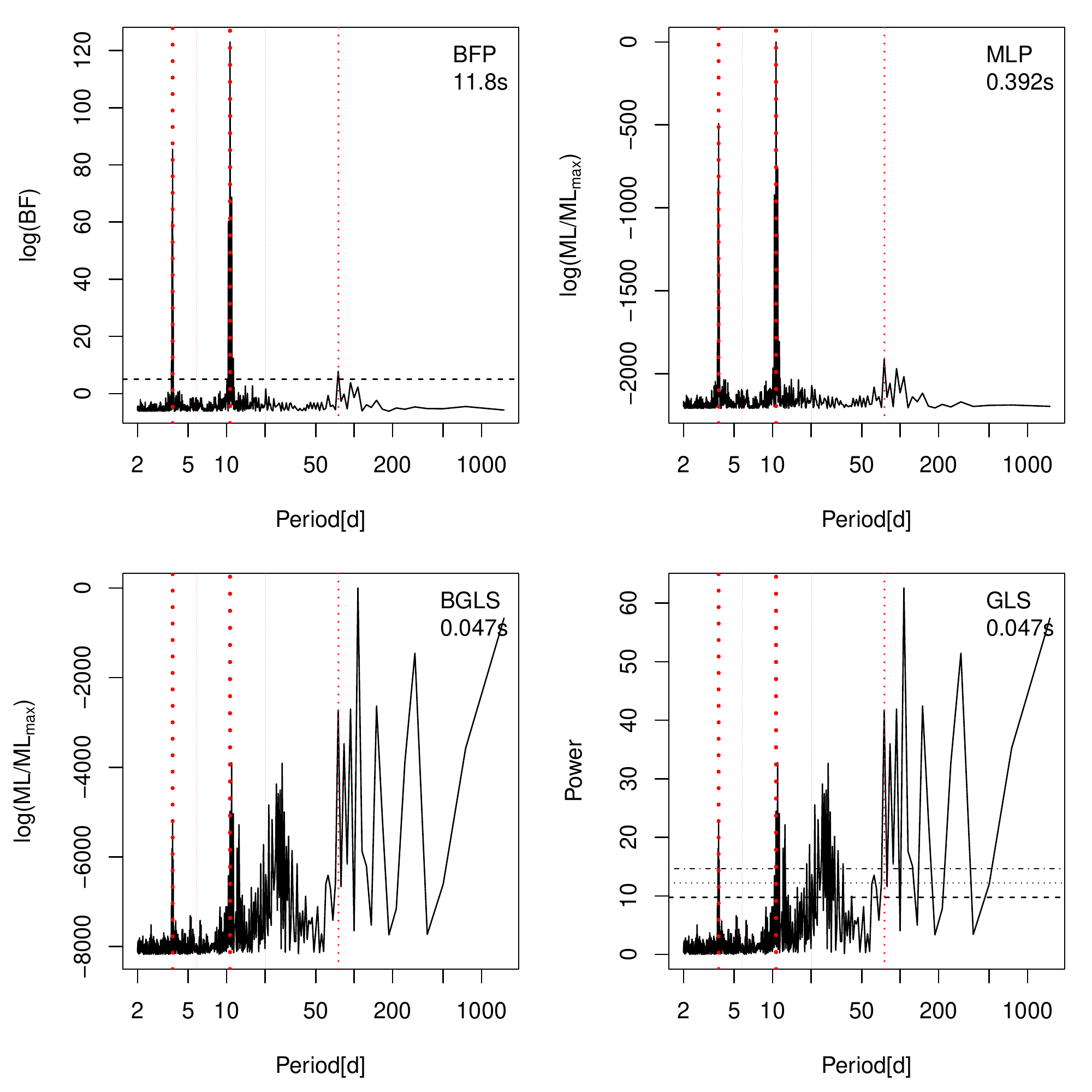}
\caption{Comparison of the periodograms of BFP, MLP, GLS and BGLS for the white noise model for the second RV challenge data set with the observational time stamps of $\tau$ Ceti. The signals are denoted by red dotted lines and the dot size is proportional to the semi-amplitude of the signal. The horizontal lines in the GLS represent the 0.001, 0.01, 0.1 false alarm probabilities. The logarithmic BF threshold of 5 is shown by the horizontal dashed line. For each periodogram, the CPU time is shown under the periodogram name in the upper right-hand corner. }
\label{fig:index}
\end{figure*}

To test this, we compare the periodograms of BFP and MLP with the Bayesian generalized Lomb-Scargle (BGLS; \citealt{mortier14}) and generalized Lomb-Scargle (GLS;  \citealt{zechmeister09}) periodograms of the second RV challenge data set in Fig. \ref{fig:index}. To account for the index-dependent noise, we linearly include all the supplied activity indexes, S-index, BIS and FWHM, in the RV model. To compare periodograms with/without accounting for indices, we use a white noise model by setting $q=0$ and $N_{\rm AP}=1$. We don't use red-noise model because we aim to see the improvement of BFP and MLP by including activity indices in the model. 

In Fig. \ref{fig:index}, we see great improvement in the signal to noise ratio of the injected signals for BFP and MLP with respect to those in the BGLS and GLS. Because the parameter of the trend component is optimized/marginalized, the BFP/MLP does not show long period powers as the BGLS and GLS. The powers corresponding to the signals become unique after accounting for the linear correlation between RVs and indices. In particular, the BF/ML for the rotation period around 25\,d is very low compared with the high probability/power in the BGLS/GLS. Although the BGLS and GLS are easy to calculate, the computation of MLP only takes 0.392\,s but greatly improves the periodogram. Nevertheless, such a huge improvement is not found for real RV data according to our analyses, indicating unrealistic or oversimplified artificial noise in the synthetic data of \cite{dumusque16a}. This is also part of the reason why we only use the white noise model in the calculation of BFP and MLP.

\begin{figure*}
\centering
\includegraphics[scale=0.6]{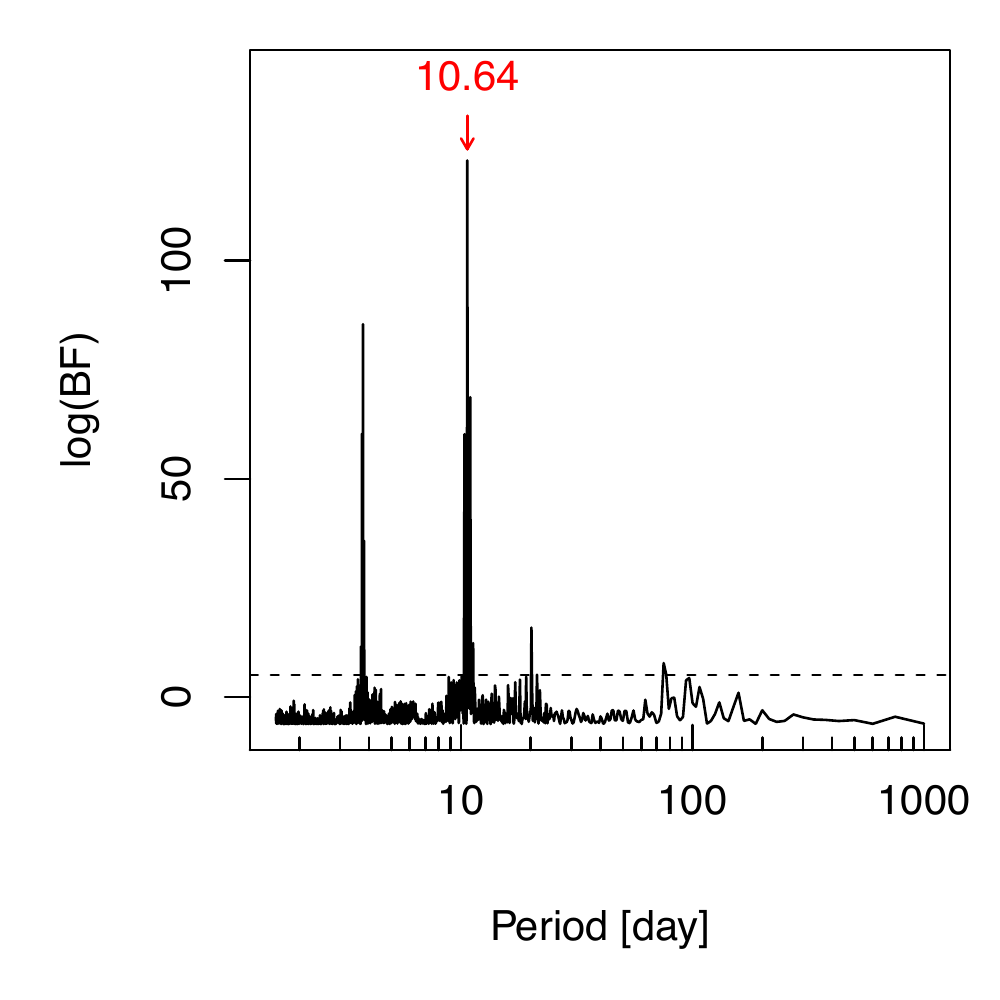}
\includegraphics[scale=0.6]{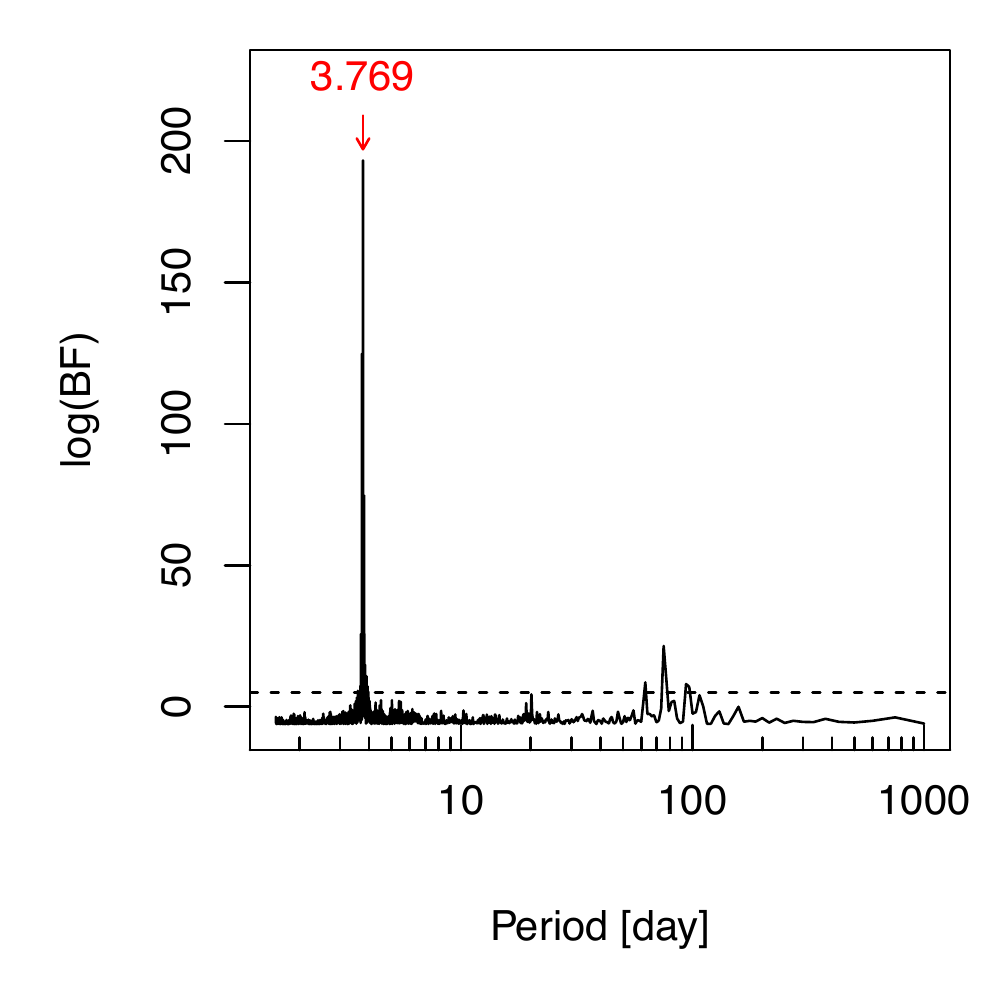}
\includegraphics[scale=0.6]{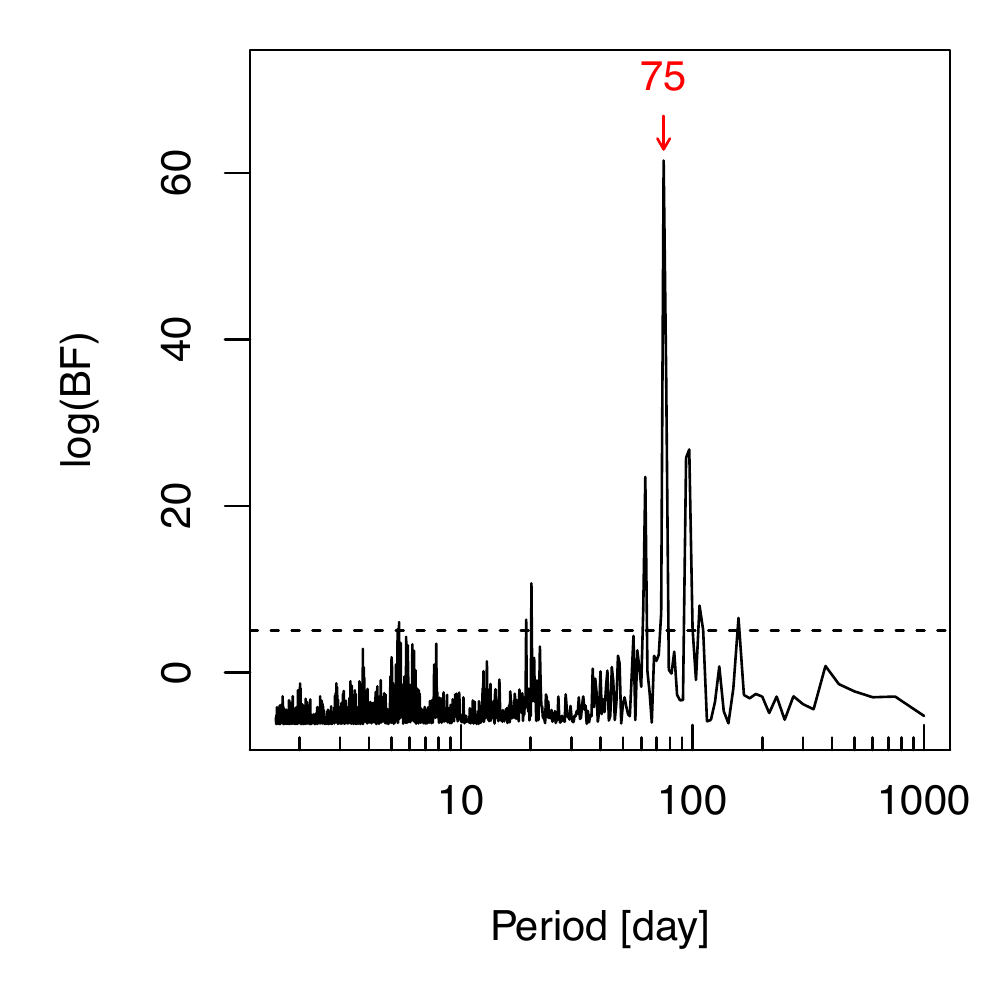}
\includegraphics[scale=0.6]{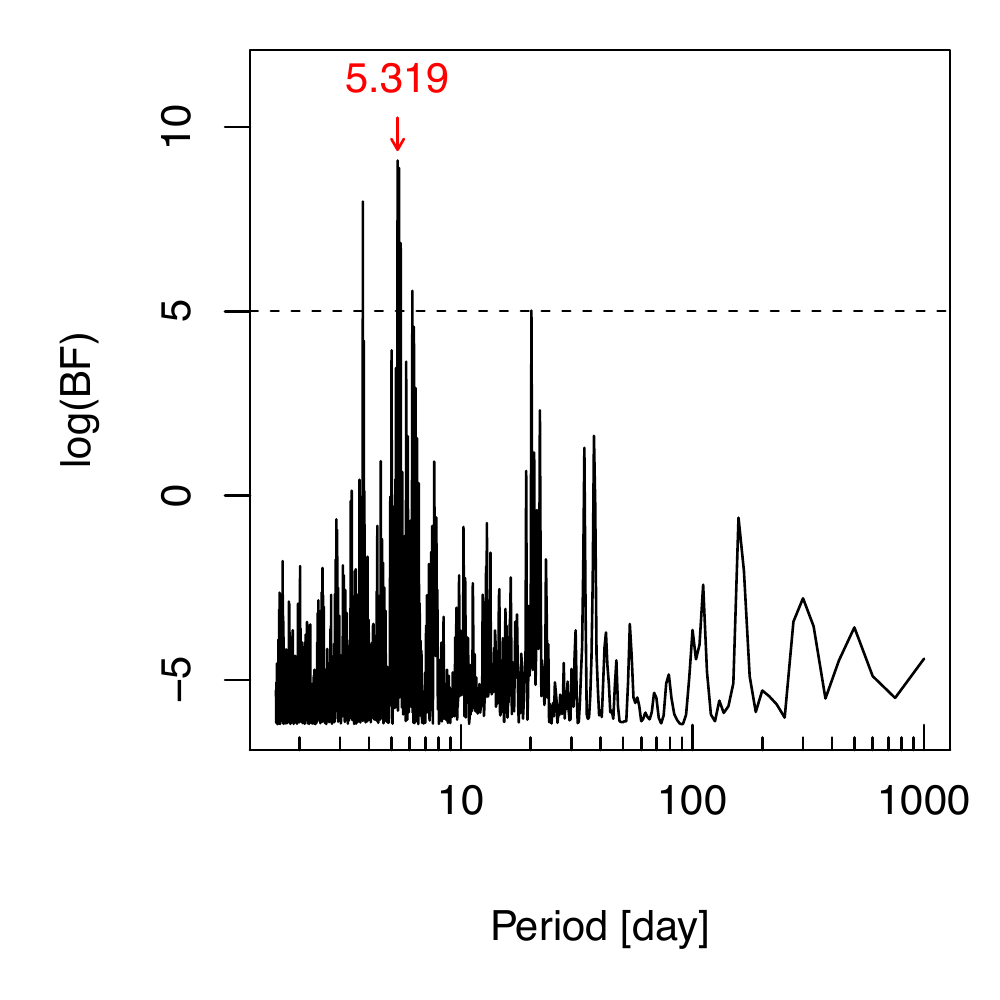}
\includegraphics[scale=0.3]{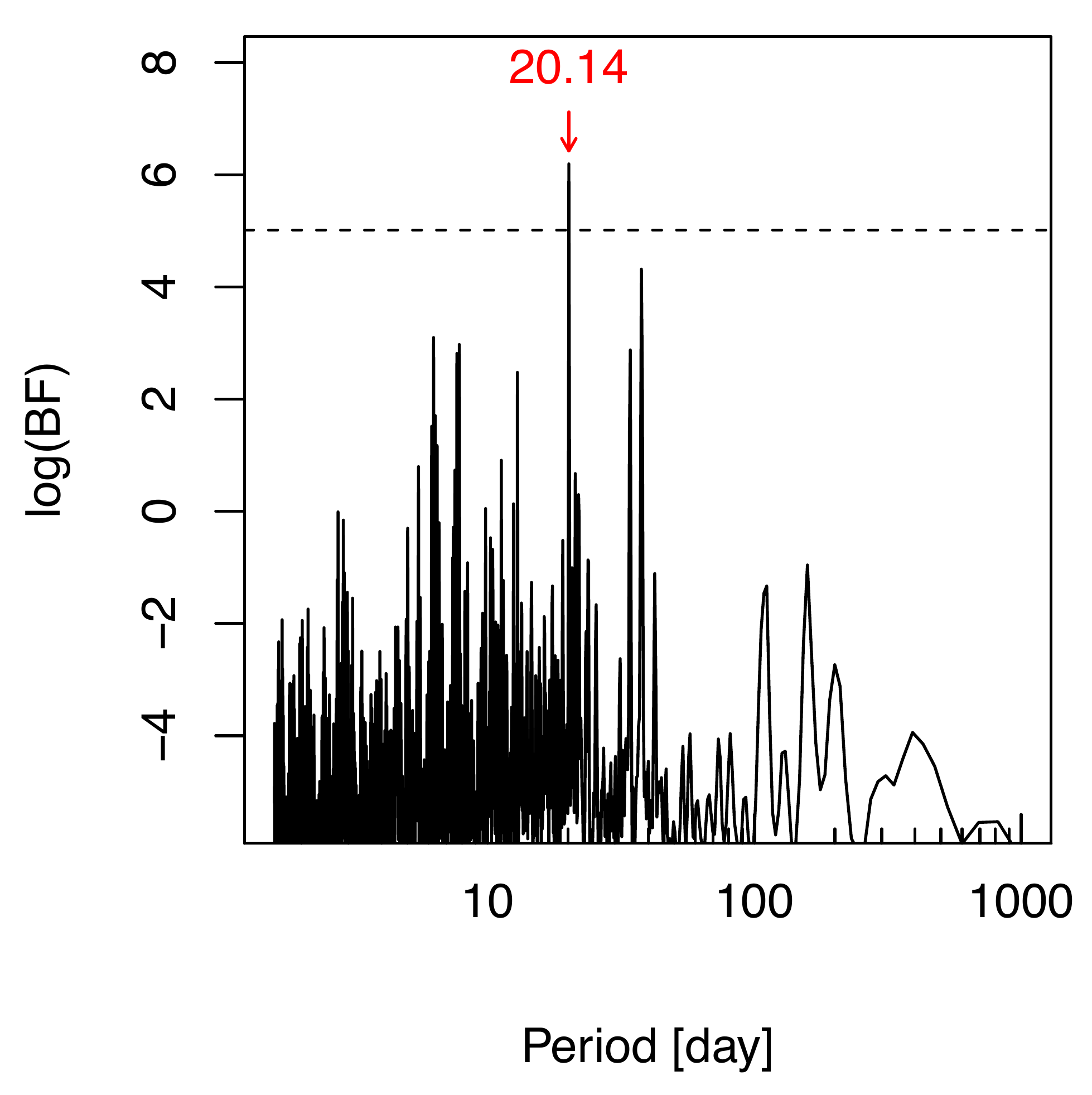}
\includegraphics[scale=0.3]{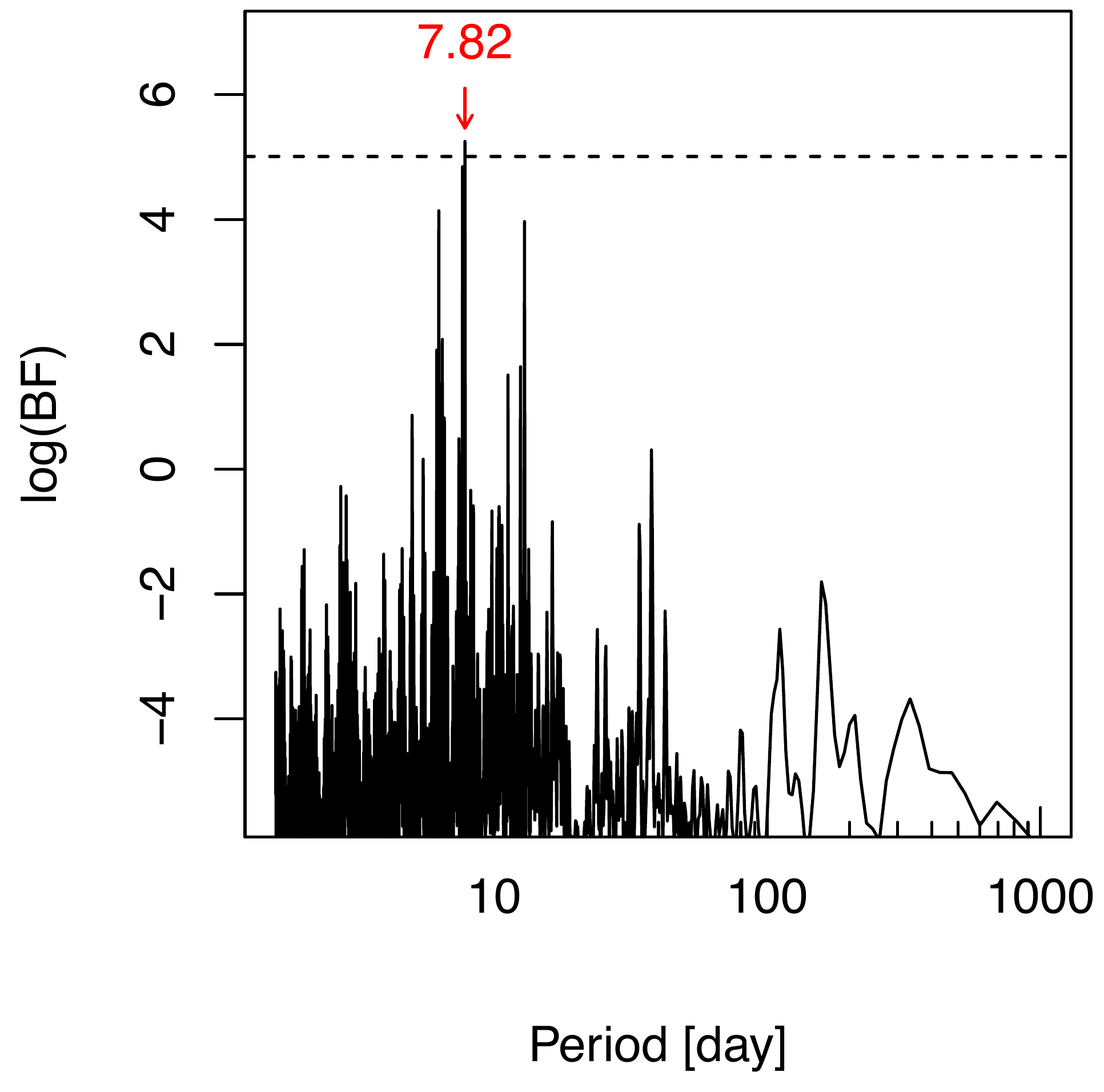}
\caption{The BFPs of the data (top left) and the residuals after subtracting the first (top right), second (bottom left), third (bottom right), circular signals. To subtract signal precisely, these BFPs are oversampled by a factor of 2 or 3.}
\label{fig:res4}
\end{figure*}

In the BFP, the strongest three injected signals at periods of 3.77, 10.64 and 75.28 d with semi-amplitudes of 2.75, 2.85 and 1.35 m/s all have logarithmic BFs larger than 5 and thus may be identified. These three signals are also the only signals recovered for this data set by the research teams in the RV challenge \citep{dumusque16b} though not by all teams. To confirm these three signals further, we subtract the signals from the data sequentially and show the residual BFPs in Fig. \ref{fig:res4}. We see that the strongest three signals are recovered while the weaker 5.79 and 20.16 d signals can be recovered to a reasonable precision. In particular, the 5.79\,d signal is not accurately recovered probably due to an incomplete subtraction of signals or an over-subtraction of the 10.64\,d signal which gives rise to a false-positive around its harmonic, 5.3\,d. Thus a detection of harmonics of a known signal  probably indicates the existence of a real signal at a similar period, which is not rare according to the distribution of the period ratio of exoplanets \citep{steffen14}. Considering this problem of signal subtractions, it should be noted that the BFP is expected to be used in combination with full Bayesian methods to detect signals and that with semi-amplitudes of 0.27 and 0.34\,m/s these signals are rather weaker than we expect to detect with confidence. Despite not ``properly'' recovering all the weaker signals in Fig. \ref{fig:res4}, it is notable that the BFP is able to recover the 1.35 m/s signal which has a $K/N$ ratio\footnote{This signal-to-noise ratio is introduced by \cite{dumusque16b} to measure the significance of a signal. The $K/N$ ratio for signal with semi-amplitude of $K$ is defined as $K/N\equiv K/RV_{\rm rms}\times \sqrt{N_{\rm obs}}$, where $RV_{\rm rms}$ is the standard deviation of RVs after removing the best-fit trend and correlation with noise proxies and $N_{\rm obs}$ is the number of observations. } of 7.6, close to the detection limit of 7.5 according to the analysis of RV challenge results \citep{dumusque16b}. Thus our recovery of simulated signals based purely on the BFP is pleasing and realistic, considering that our team has detected signals with $K/N$ as low as 5 without announcing false positives in the RV challenge competition \citep{dumusque16b}.


\subsection{Periodograms for time-correlated noise}\label{sec:ma}
As concluded in \cite{baluev13}, accounting for red noise in the RV time series is crucial for correctly identifying Keplerian signals. However, the periodograms used by most researchers in the community are based on the implicit assumption that the noise is white. To overcome this problem, Bayesian methods implemented by various algorithms have been developed to properly model the time and wavelength-correlated noise (e.g. \citealt{ford07}, \citealt{tuomi12} and F17a). Recently a framework of Gaussian process is developed to mitigate the red noise caused by stellar activity \citep{rajpaul15}. But the Gaussian process is probably too flexible to be the Goldilocks noise model, which avoids both false positives and negatives \citep{feng16}. F17a have demonstrated this by comparing different MA models in the Bayesian framework in order to select the Goldilocks noise model.

\begin{figure*}
\centering
\includegraphics[scale=0.6]{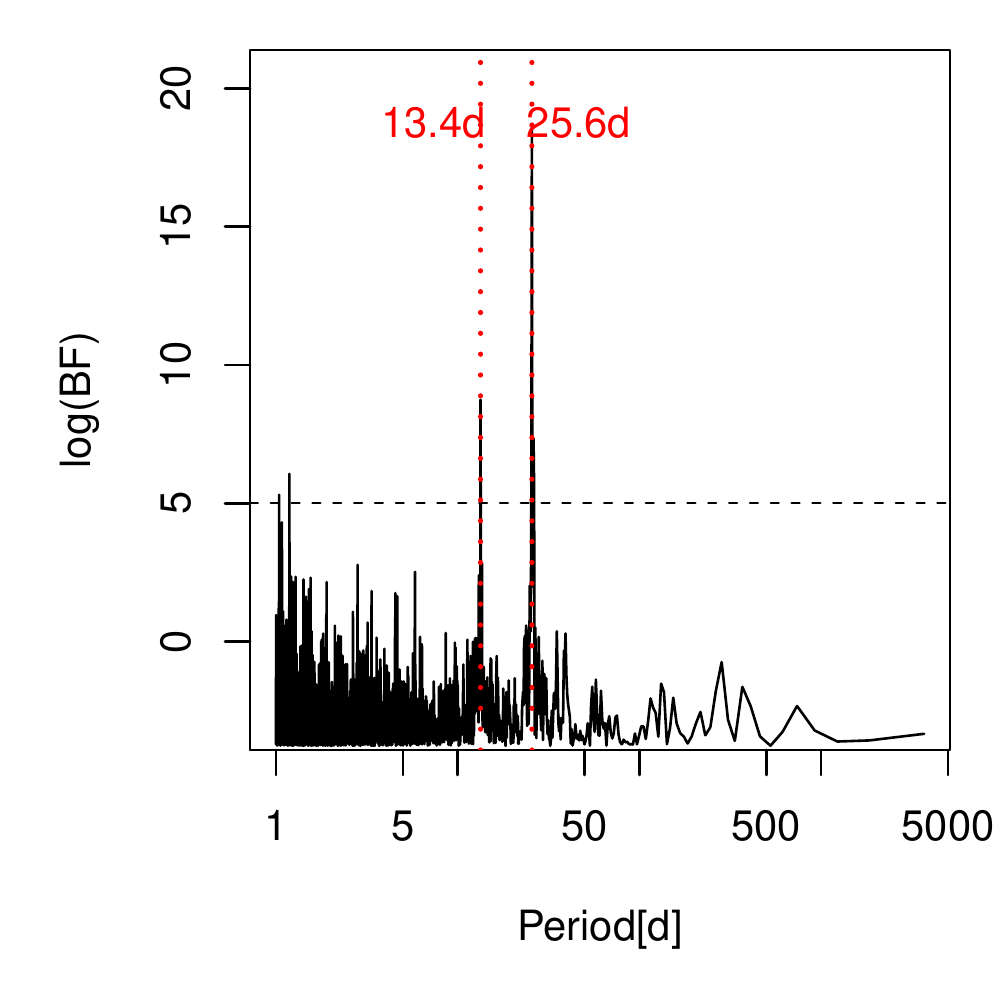}
\caption{Similar to Fig. \ref{fig:index}, but for the first order MA model combined with the activity indices for the HARPS data set of HD41248. The blue dotted line denotes the 18.4\,d planetary candidate reported by \protect\cite{jenkins13} based on Bayesian analysis of the same data set. The red dotted lines denote the signal identified by the BFP.}
\label{fig:red}
\end{figure*}

However, the Bayesian approach is computationally expensive due to the requirement of intensive sampling of the posterior density and computations of integrated likelihoods to estimate Bayes factors. To efficiently account for the red noise as well as to visualize the periodic signals, we use the BFP/MLP to account for red noise in the RV data. In section \ref{sec:comparison}, we show that the noise model with one MA component and without differential RVs is favored by the HARPS data of HD41248. With this noise model, we calculate the BFP and MLP, and compare them with the GLS and BGLS in Fig. \ref{fig:red}. 

In this figure, we observe that the $\sim$18.4\,d signal detected by \cite{jenkins13} and \cite{jenkins14} is not as significant as a 13.4\,d signal in the BFP. The long-period signals appearing in the BGLS, GLS and MLP are strongly weakened in the BFP. We also see that the power of the 26\,d signal in the BFP is stronger than that in the other periodograms probably because the red noise in the data would broaden the BF distribution around the signal if it were not accounted for. Although the model used in the calculation of BFP is similar to that used by \cite{jenkins14}, our results are not sensitive to the choice of noise models since similar signals are identified in different periodograms and are also independently detected by \cite{santos14}. 

\begin{figure*}
\centering
\includegraphics[scale=0.55]{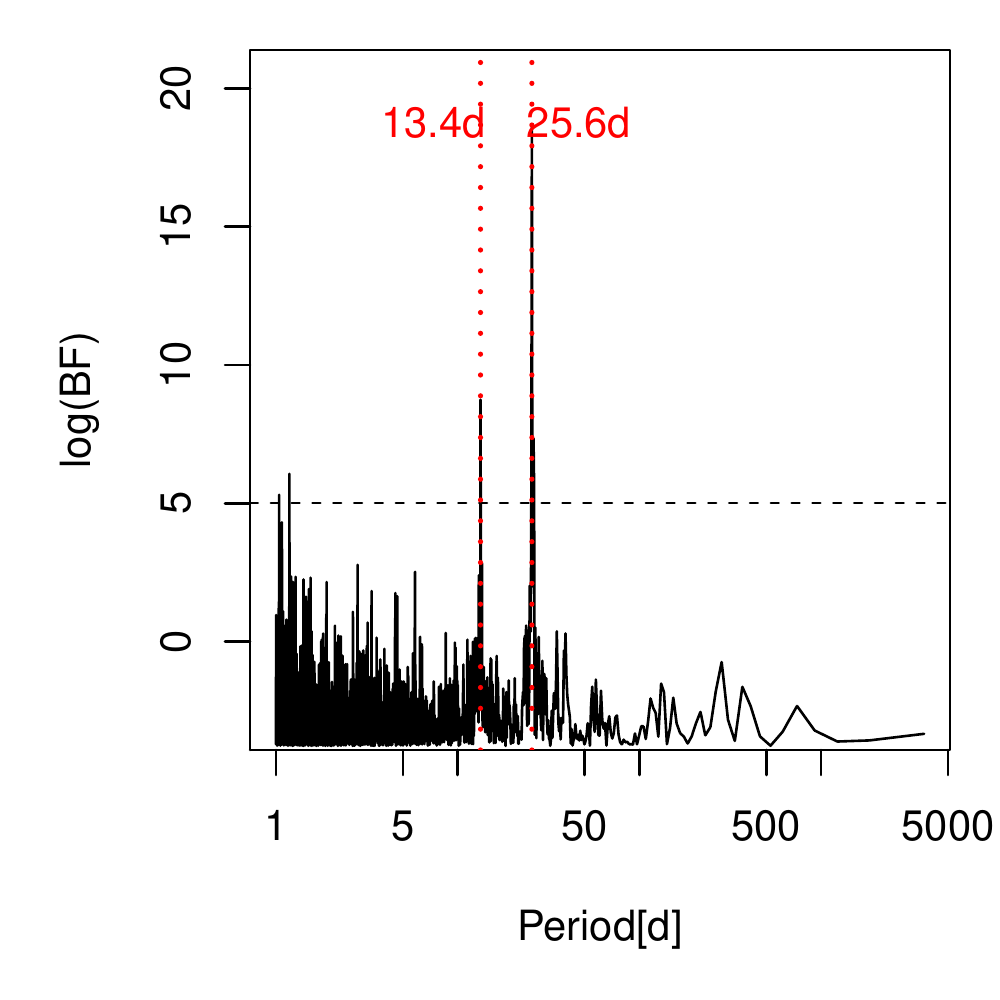}
\includegraphics[scale=0.55]{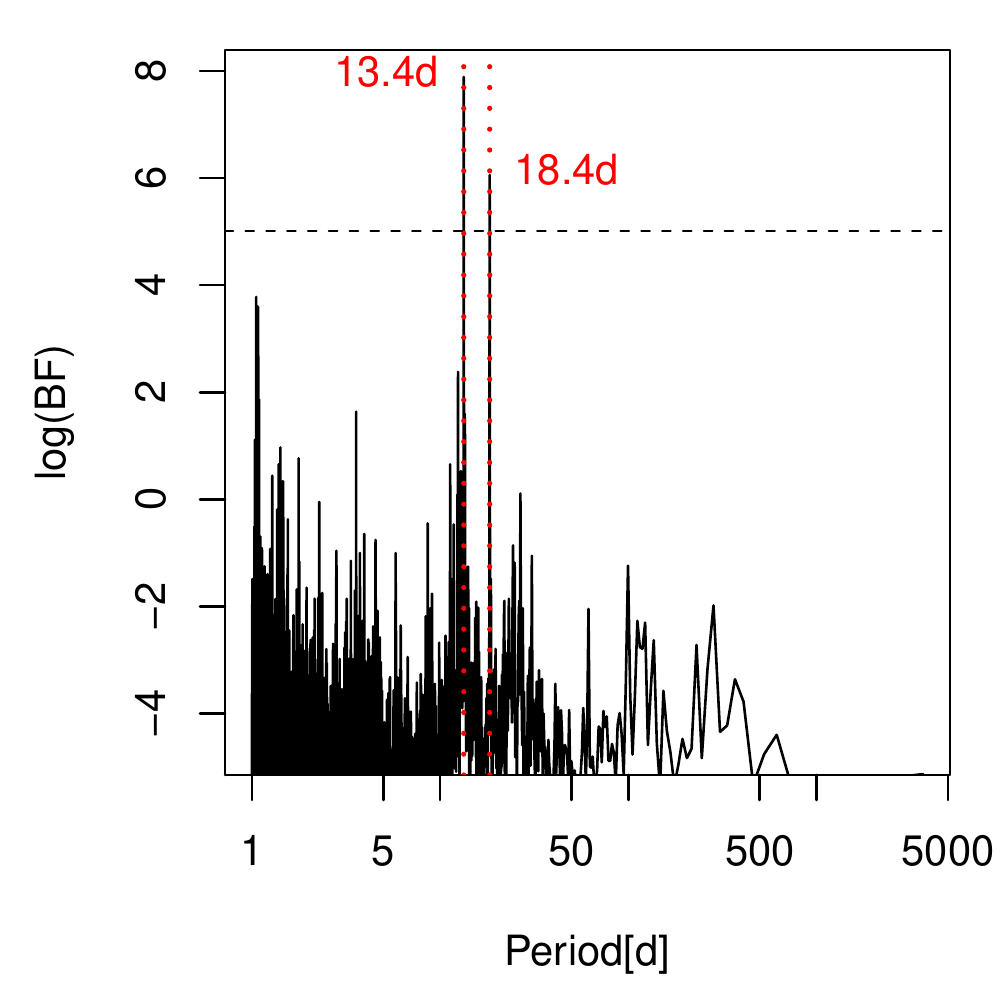}
\includegraphics[scale=0.55]{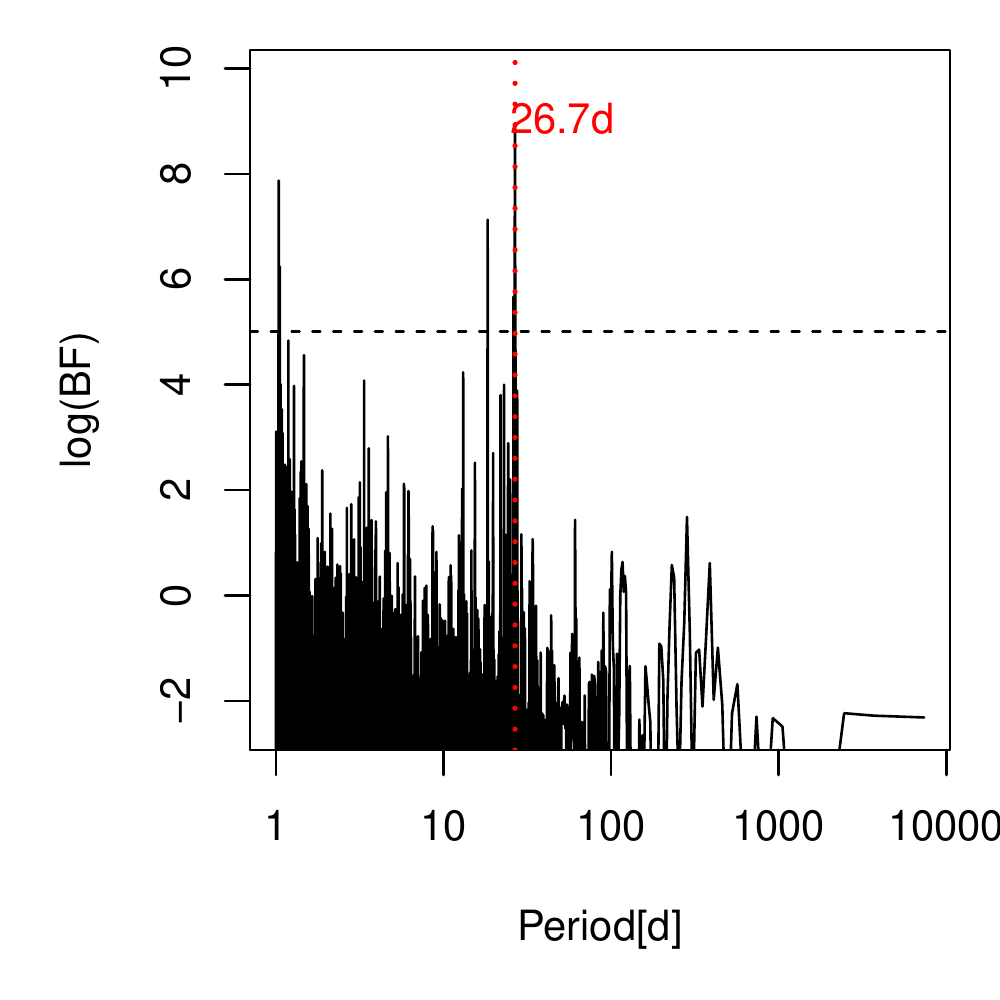}
\includegraphics[scale=0.55]{per221_MLPsig2.pdf}
\includegraphics[scale=0.55]{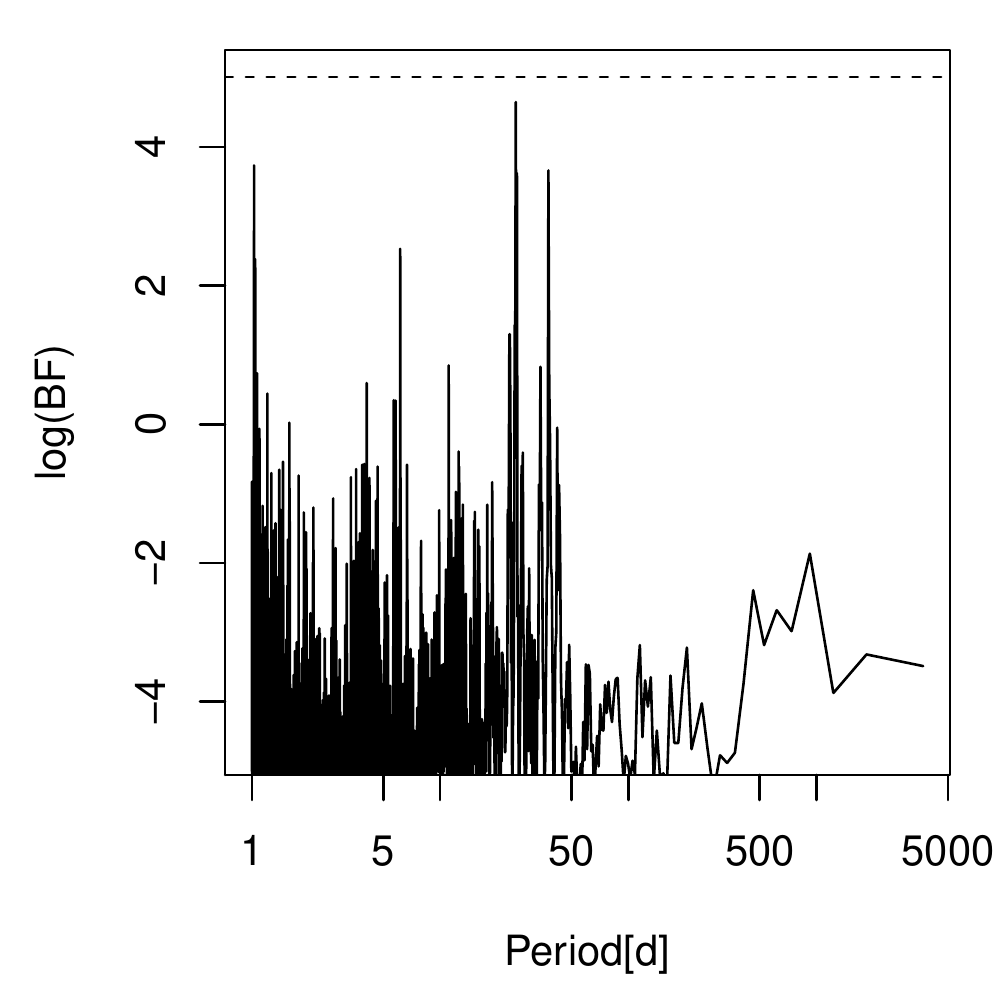}
\caption{The BFPs for the residuals after subtracting the 13.4\,d signal (top left), the 26\,d signal (top middle), the 13.4 and 26\,d signals (top right), the circular 18.4\,d signal (bottom left), the 13.4, 26 and 290.1\,d signals (bottom right) from the HARPS data of HD41248. }
\label{fig:HD41248res}
\end{figure*}

To find and compare signals, we subtract signals quantified using the AM posterior sampling from the data, and show the residual BFP in Fig. \ref{fig:HD41248res}. We see in the top left panel that the residuals strongly support the existence of the 26\,d signal in the data subtracted by the 13.4\,d signal. If we subtract the 26\,d signal from the data, the 13.4\,d one does not disappear, indicating that these two signals are independent signals rather than harmonics of each other, as interpreted by \cite{santos14}. We further subtract the 13.4 and 26\,d signals from the data, and find a signal at a period of about 26.7\,d in the residual BFP (see the top right panel in Fig. \ref{fig:HD41248res}). This signal is probably caused by an incomplete subtraction of the 26\,d signal which corresponds to a rather broad peak in the GLS shown in Fig. \ref{fig:red}. This indicates a contribution from stellar noise to this signal, as suggested by previous analyses \citep{santos14,jenkins14}.

To compare the 13.4 and 18.4\,d signals, we subtract the circular 18.4\,d signal from the data because no MCMC chains has identified this signal for the one-planet model. We show the residual BFP in the bottom left panel. We see high BF around 13.4\,d, indicating that the 18.4\,d signal is probably an alias of the 13.4\,d signal. Although we do not find strong signals other than the $\sim$26\,d one in the top right panel, we find the third signal at a period of about 290\,d using the AM posterior sampling. It increases the logarithmic BF by a factor of 2.5 with respect to the two-planet model, despite a failure in passing the logarithmic BF threshold. However, this signal may be connected to the $\sim$26 and 13.4\,d signals since $1/(2/25.6-1/13.4)=1/285.9$. The 290\,d signal together with the 13.4 and 26\,d signals are subtracted from the data to calculate the residual BFP shown in the bottom right panel of Fig. \ref{fig:HD41248res}. This residual BFP does not show any significant signals. Thus there are at most three signals at periods of 13.4, 26 and 290\,d in this data set. To study the nature of these signals, we will test their consistency in time in section \ref{sec:mp}.

\subsection{Periodograms for noise correlated in time and wavelength}\label{sec:dRV}
The noise in RVs is caused not only by stellar activity which is partially recorded by activity indices but also by wavelength-dependent atmospheric and instrumental effects (F17a). Previous periodograms do not take the wavelength-dependent noise into account and thus are biased in this respect in terms of identifying potential Keplerian signals. By accounting for differential RVs as additional noise proxies, the BFP and MLP are able to remove the wavelength-dependent noise to a large extent.

\begin{figure*}
\centering  
\includegraphics[scale=0.65]{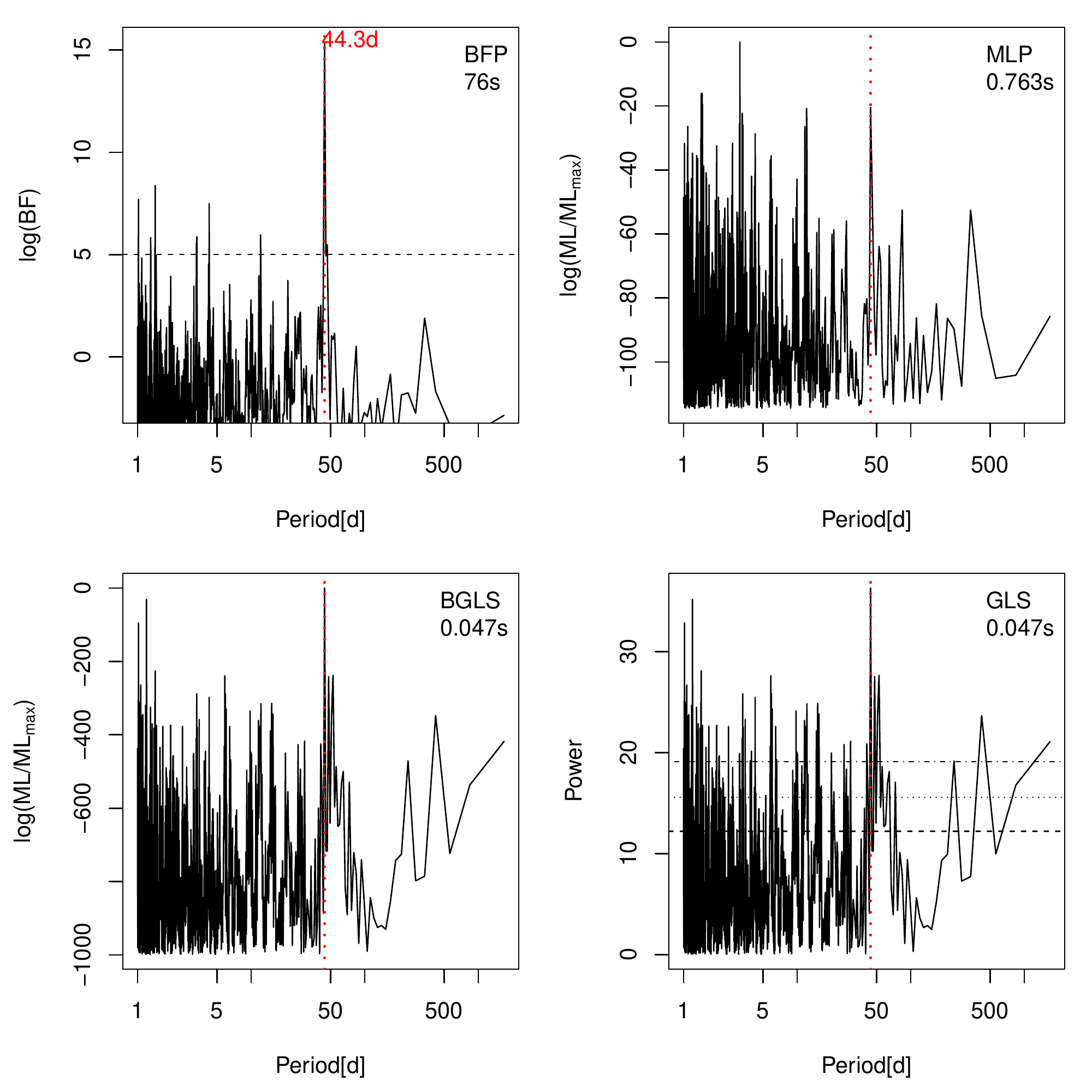}
\caption{Similar to Fig. \ref{fig:index}, but for the model including differential RVs for the HARPS data set of HD177565.}
\label{fig:dRV}
\end{figure*}

To test this, we compare the BFP, MLP, BGLS and GLS for the HARPS RV data set of HD177565. We adopt the noise model including linear functions of activity indices and the 3AP differential RVs. This noise model is favored by the data based on the BFs calculated both by the AM method and by the BFP (see Table \ref{tab:BFs}). Adopting this noise model, we calculate the BFP, MLP and other periodograms, and show them in Fig. \ref{fig:dRV}. Compared with the other periodograms, 44\,d appears to be more prominent in the BFP probably due to the accounting for correlated noise, as shown in Fig. \ref{fig:red}. The 44\,d signal is not significant in the MLP, indicating a bias introduced by noise subtraction which is mentioned in section \ref{sec:MLP} and \ref{sec:index}. 

\begin{figure*}
\centering  
\includegraphics[scale=0.55]{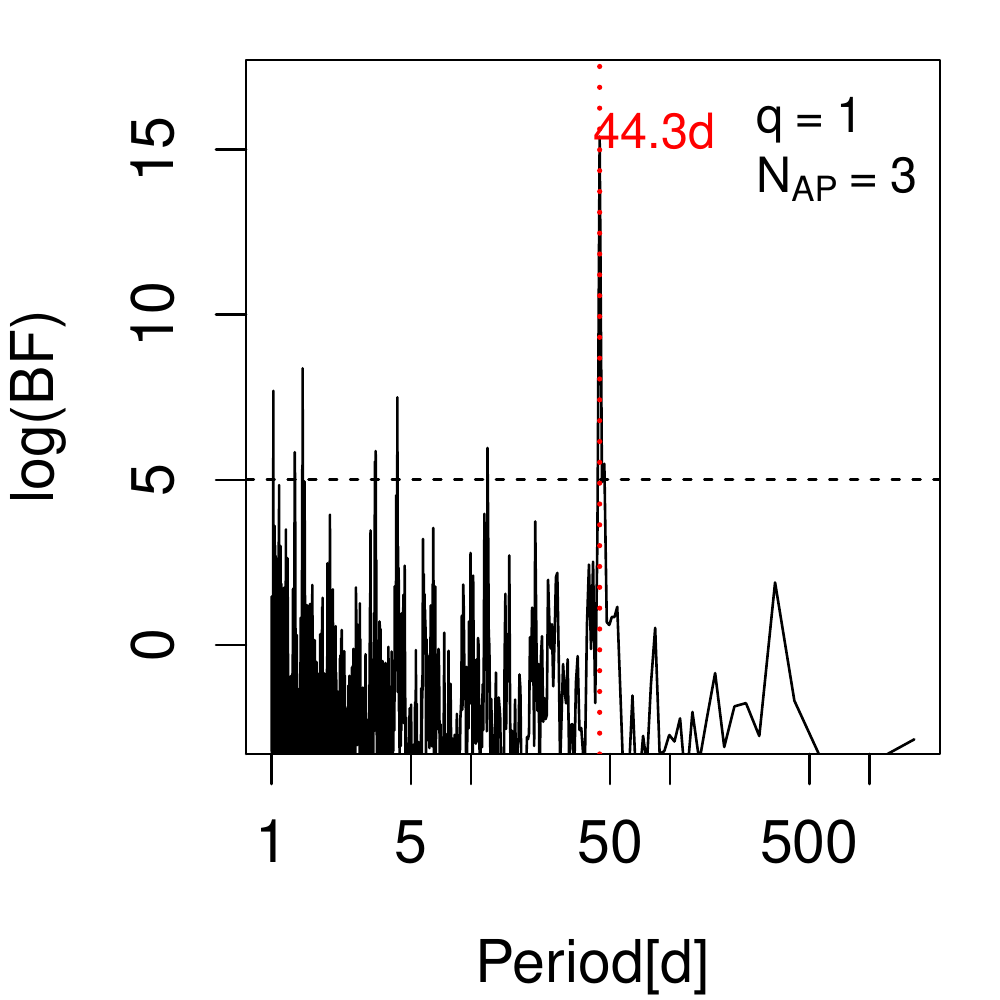}
\includegraphics[scale=0.55]{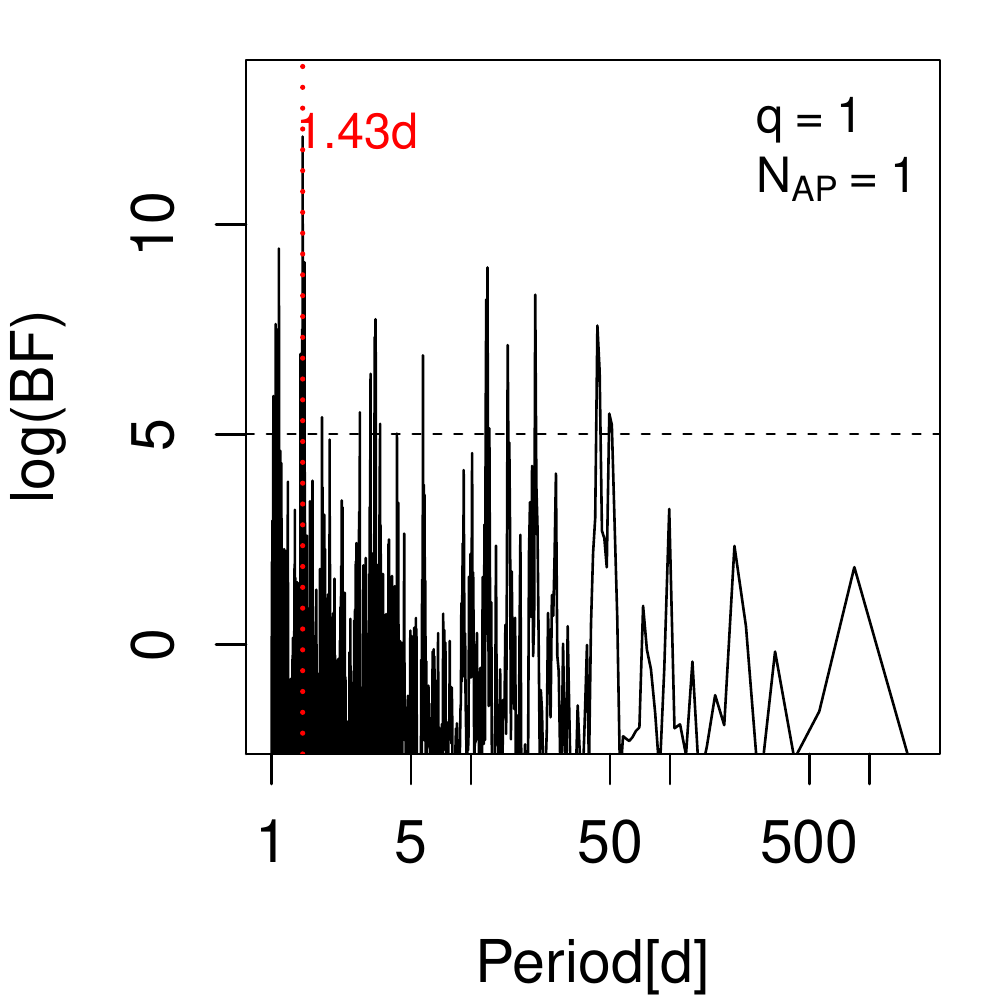}
\includegraphics[scale=0.55]{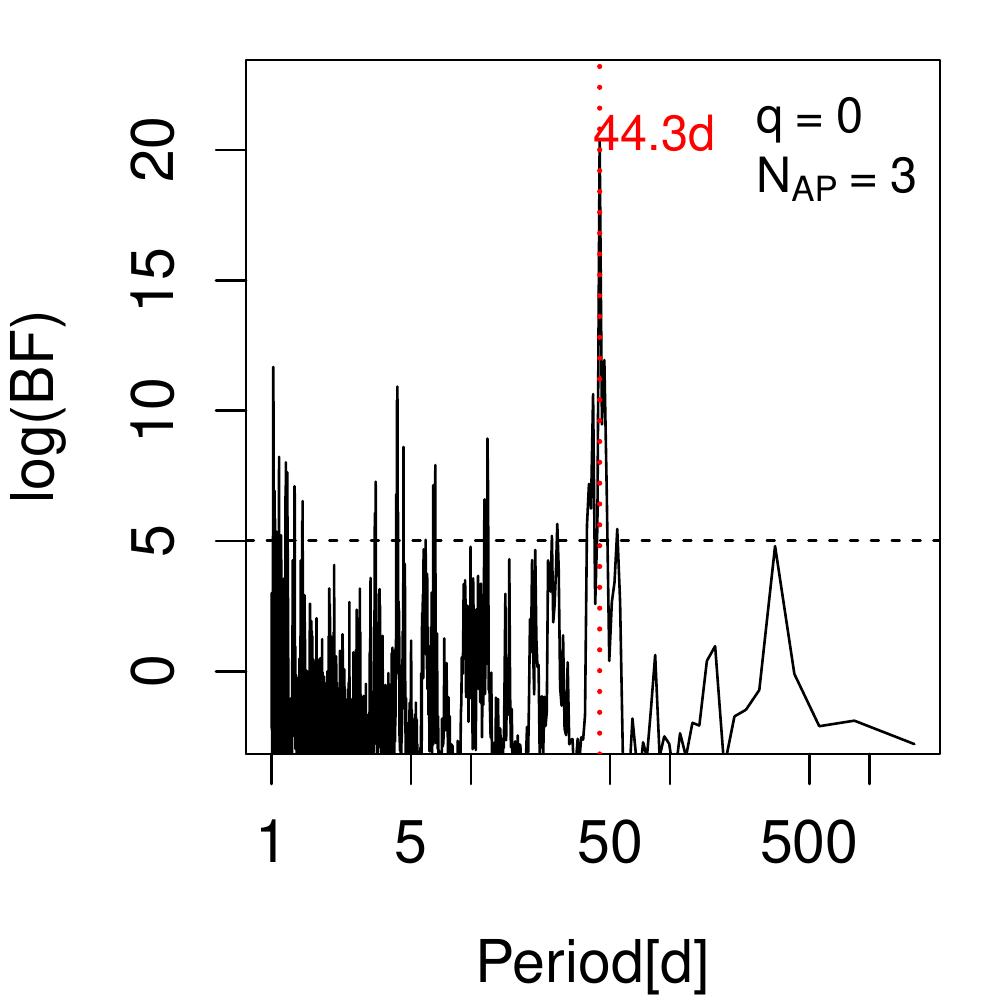}
\caption{The BFPs for models with $\{q,N_{\rm AP}\}=\{1,3\}$, $\{1,1\}$ and $\{0,3\}$ for HARPS data of HD177565. The signals with maximum BFs are denoted with red dotted lines. }
\label{fig:3models}
\end{figure*}

To illustrate the relative roles of using an MA model and accounting for differential RVs in reducing noise, we calculate the BFPs for models with $\{q,N_{\rm AP}\}=\{0,3\}$ and $\{1,1\}$, and compare them to the previous BFP, top-left in Fig. \ref{fig:3models}. We observe that the differential RVs play an important role in reducing the wavelength-dependent noise and improve the significance of the $\sim$44\,d signal. Without dependence on the differential RVs, the BFP would identify a signal at a period of 1.43\,d, which is much weaker than the 44\,d signal based on the AM-based samplings. On the other hand, the MA model plays a role in reducing the time-correlated noise and the false positive rate. However, based on the the difference in the BF ranges of the BFPs, setting $q=1$ would appear to weaken the signal somewhat presumably from part of the periodic variability being interpreted as red noise. Thus as suggested by \cite{feng16}, the MA model together with other stochastic models should be penalized more than the noise proxies for this noise model comparison. We use the first order MA model combined with 3AP differential RVs to model the RV noise of HD177565. 

\begin{figure}
\centering  
\includegraphics[scale=0.8]{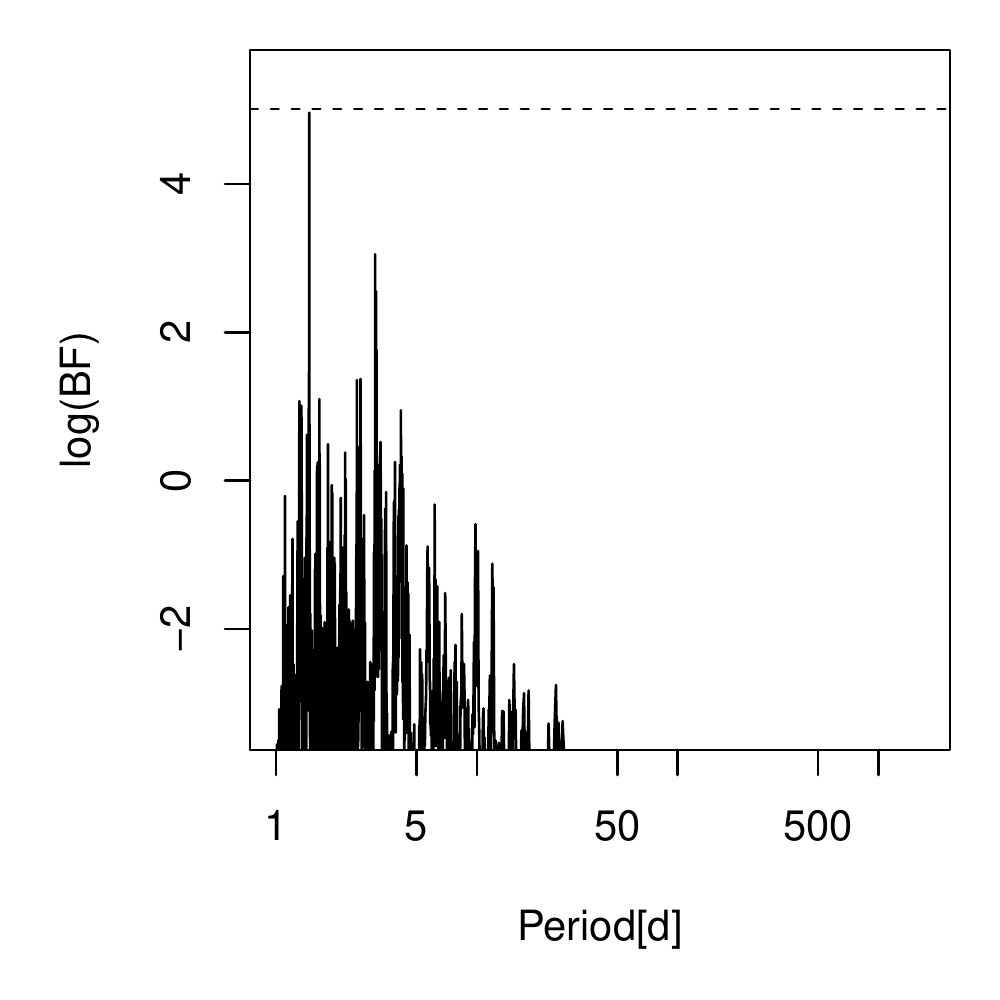}
\caption{The BFP for the HARPS data of HD177565 with the 44\,d signal subtracted and quantified using the AM posterior sampling. }
\label{fig:HD177565res}
\end{figure}

To find additional signals, we estimated the parameters for the 44\,d signal using AM-based posterior samplings, and subtract the optimal Keplerian component from the data. Then, we calculate the residual BFP and show it in Fig. \ref{fig:HD177565res}. Based on this residual periodogram, we cannot identify any additional significant signals. 

\begin{figure}
\centering  
\includegraphics[scale=0.7]{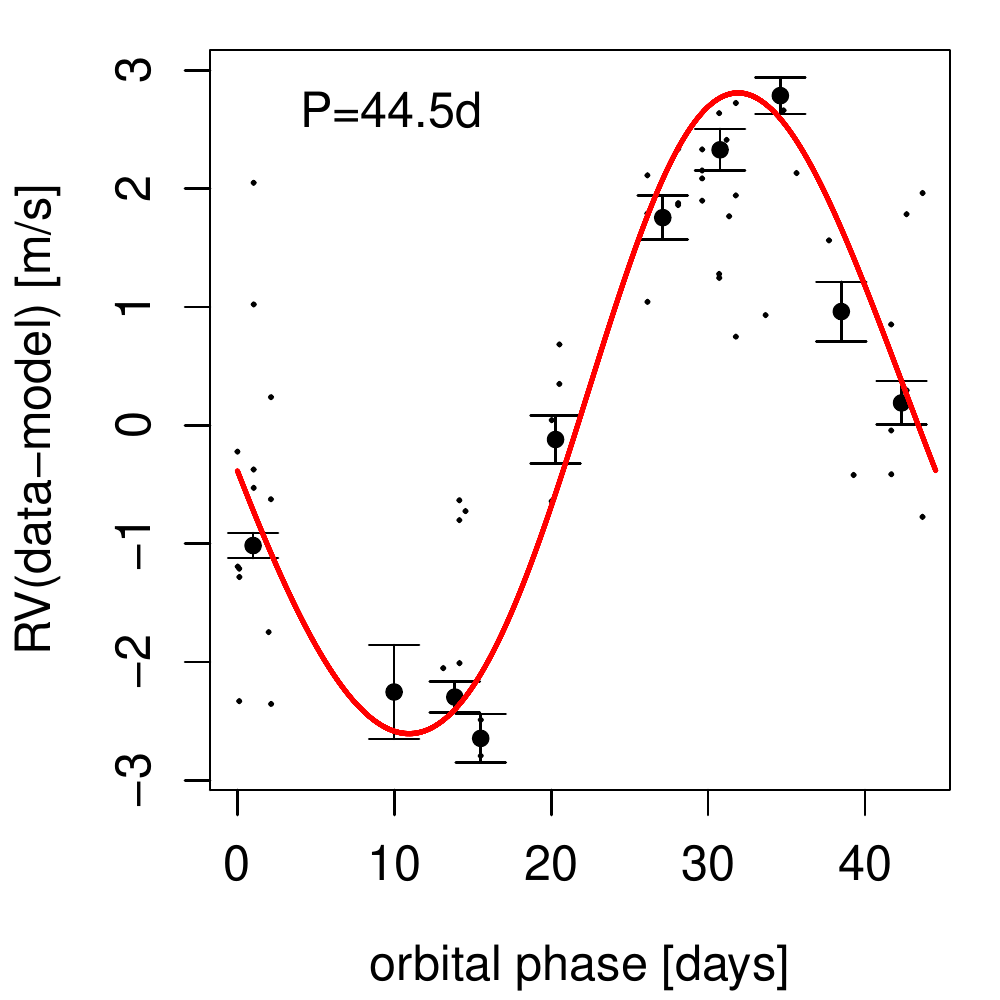}
\caption{The phase-folded data and the one-planet model prediction for the TERRA-reduced HARPS data of HD177565. The small dots represent raw RVs while the error bars denote the binned data. The noise component is subtracted from the data and the model.}
\label{fig:HD177565phase}
\end{figure}

Based on the AM-based posterior samplings, we show the phase-folded data and model predictions in Fig. \ref{fig:HD177565phase}. We observe a reasonable phase coverage for this signal, which corresponds to a planet candidate at a period of about 44\,d orbiting HD177565 on a nearly circular orbit. 
Adopting a stellar mass of 1.0\,$M_\odot$ \citep{silva12} for HD177565, we further calculate and report the parameters of this signal by tabulating maximum {\it a posteriori} (MAP) estimates in Table \ref{tab:HD177565sig}. If this signal is caused by a planet, it has a minimum mass of 15.1\,$M_\oplus$ and semi-major axis of 0.25\,AU, and thus is a hot Neptune. Notably, there is also a debris disk orbiting HD177565 \citep{beichman06}, probably leading to a high impact rate on this planet. 
\begin{table}
  \centering
  \caption{The MAP estimation of the parameters for two signals detected in the TERRA-reduced HARPS data of HD177565. The uncertainties of parameters are represented by the values determined at 1\% and 99\% of the cumulative posterior density. We estimate the minimum planetary mass and semi-major axis using a stellar mass of 1.0\,$M_\odot$ \citep{silva12} with an assigned $1\sigma$ uncertainty of 0.1\,$M_\odot$.}
\label{tab:HD177565sig}
  \begin{tabular}  {c*{5}{c}}
\hline 
Parameters& HD 177565b\\\hline
$P$\,(d)&44.505 [44.212, 45.091]\\
$K$\,(m/s)&2.71 [1.72, 3.83]\\
$e$&0.0593 [0.00185, 0.231]\\
$\omega$\,(rad)&5.41 [0.0679, 6.21] \\
$M_0$\,(rad)&2.57 [0.0686, 6.23]\\
$m\sin{i}$\,($M_\oplus$)&15.1 [9.05, 21.5]\\
$a$\,(au)&0.246 [0.227, 0.265]\\\hline
  \end{tabular}
\end{table}

\subsection{Moving periodogram}\label{sec:mp}
If the semi-amplitude of a periodic signal does not change over time, the signal should be consistently identified in different data chunks which are data subsets taken at different time intervals. One method for this consistency test is to conduct Bayesian analyses of different data chunks. But this is computationally expensive, and cannot necessarily be applied because of uneven data sampling. An alternative method is to calculate the periodogram of the data within a time window. We move the window with small time steps, and repeat the calculation of periodograms until the whole time span is covered. The assembly of these periodograms forms a 2D periodogram map. This is the so-called ``moving periodogram'', which has been used in the analysis of Kepler light curves by \cite{ramsay16} and is introduced in the analysis of RV data here. To make the moving periodogram computationally efficient, we calculate the MLP for each data chunk by subtracting noise component from the data.  

\begin{figure*}
  \centering
  \hspace*{-0.3in}
\includegraphics[scale=0.65]{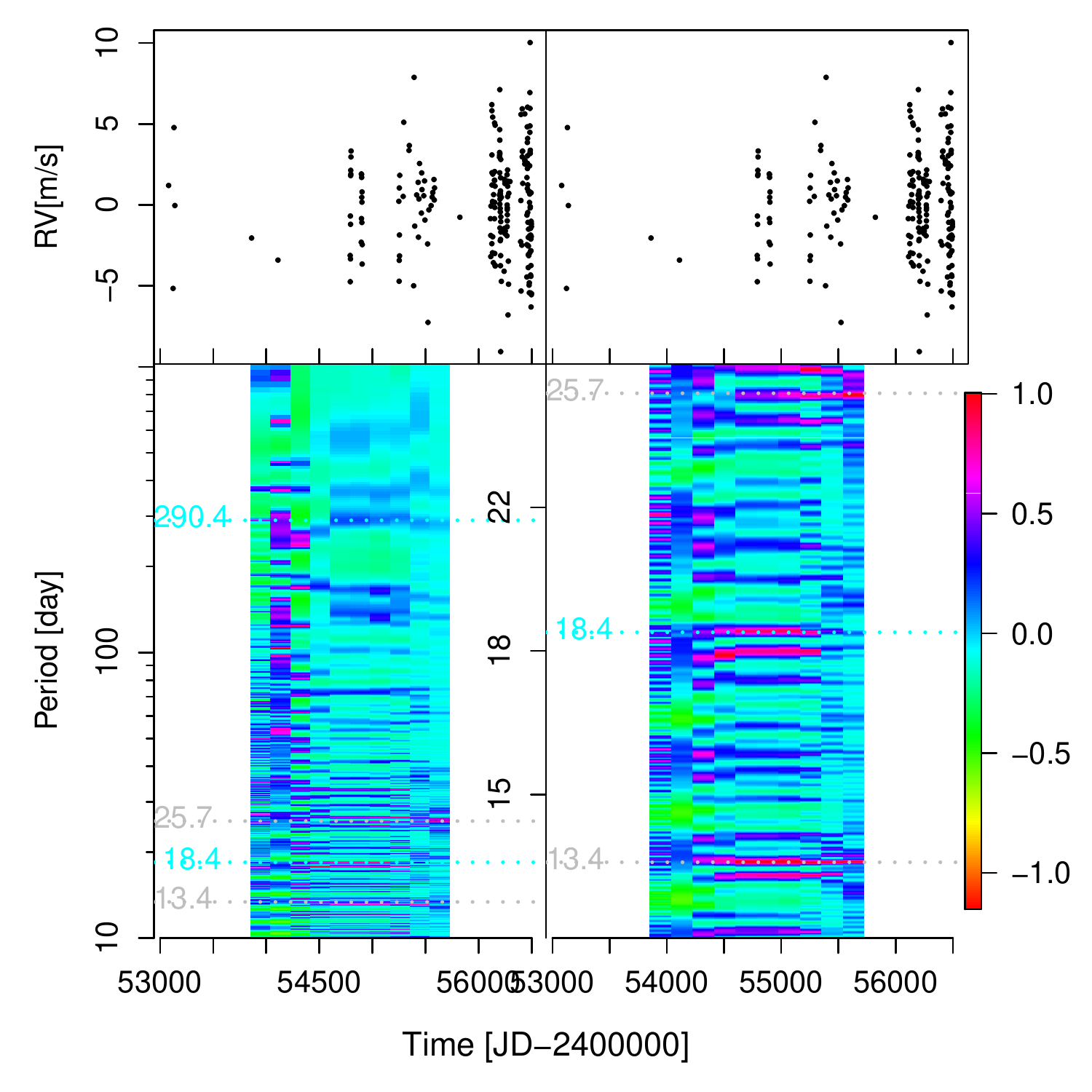}
\caption{The MLP-based moving periodogram of the HARPS RV data set of HD41248. The periodogram is calculated in a 2000\,d moving time window covering the whole time span in 10 steps. The periodogram powers for each step is encoded by colors and shown vertically at the center of the time window. Thus the time span of the moving periodogram is shorter than the data time span by 2000\,d. The top panels show the noise-subtracted data while the bottom ones show the moving periodogram for period ranging from 10 to 1000\,d (left) and for a narrow period range around the short period signals (right). The color bar shows the RMLs of the MLP. To optimize the visualization of signals, the RMLs are truncated to ${\rm med(RML)}-5\sigma_{\rm RML}$, where ${\rm med(RML)}$ and $\sigma_{\rm RML}$ are the median and standard deviation of RMLs, respectively. The signals are denoted by the grey dotted lines while the signals not identified by the model are denoted by cyan dotted lines. }
\label{fig:mp}
\end{figure*}

For example, we identify two signals at periods of 13.4 and 26\,d in the HARPS data of HD41248. We remove the correlated noise component in the 2-planet model from the data and calculate the MLP with a white noise model (or set ${\bf m=d=0}$). We calculate the MLP for the data in a 2000\,d time window, and move this window to cover the whole time span within 100 steps. We calculate the MLP for each time step, and scale the MLP power to be ${\rm RML}\equiv ({\rm ML}-\overline{\rm ML})/({\rm ML}_{\rm max}-\overline{\rm ML})$ where $\overline{\rm ML}$ and ${\rm ML}_{\rm max}$ are the mean and maximum marginalized likelihood (ML). Then we use colors to encode the RML to calculate the moving periodogram, which is shown in Fig. \ref{fig:mp}. This periodogram is aimed at visualizing the consistency of signals in time rather than assessing the significance of signals. We can see that the short period signals are consistently identified over the whole time range while the $\sim$290 d signal is visible over about 1000 days centering around JD2455000 and is relatively less distinct at other times.

\begin{figure*}
\centering
\includegraphics[scale=0.55]{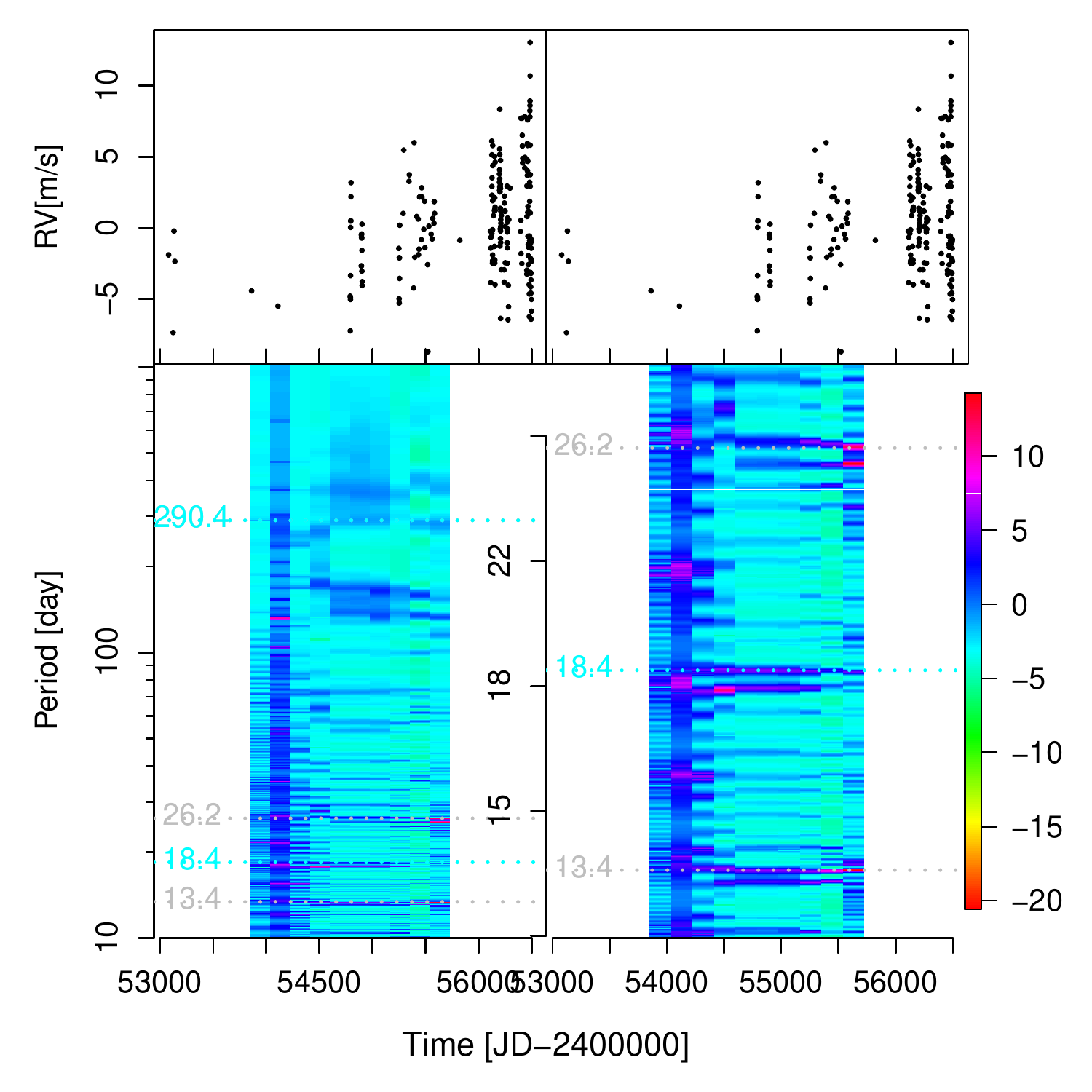}
\includegraphics[scale=0.55]{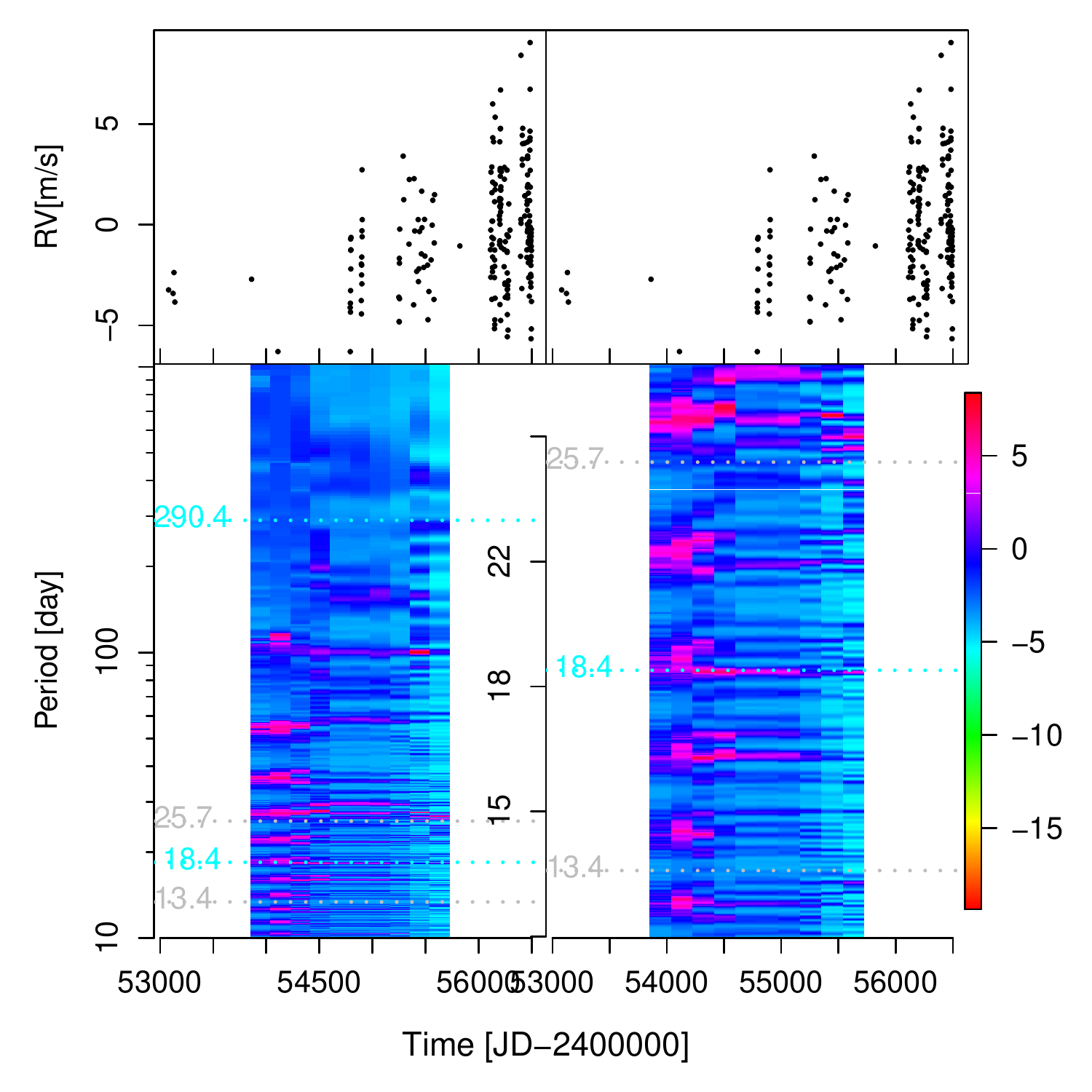}
\caption{The BFP-based moving periodograms of the HARPS RV data set of HD41248 (left) and the residuals after subtracting the 13.4 and 26\,d signals (right). The color encodes the logarithmic BFs. The logarithmic BFs are truncated to optimize the visualization of signals. The other elements are similar to Fig. \ref{fig:mp}.}
\label{fig:BFP-MP}
\end{figure*}

Instead of subtracting the correlated noise from the data, we also calculate the BFP-based periodogram of the original data to account for the possible time-varying noise properties noticed by \cite{santos14} and \cite{jenkins14}. Unlike the MLP-based moving periodogram, the BFP-based one can assess the significance of signals by encoding the logarithmic BFs with colors. Similar to the MLP-based moving periodogram, the BFP-based moving periodogram is generated from a sequence of BFPs made within a 2000\,d moving window. Since the BFP is computationally more expensive than the MLP-based one, we only cover the whole data with 10 steps. We show this moving periodogram in the left panel of Fig. \ref{fig:BFP-MP}. The BFP shows a consistent signal at a period of about 13.4\,d. Although the 18.4\,d signal also consistently appears in the BFP as noticed by \cite{jenkins14}, it is not as strong as the 13.4\,d signal and the subtraction of the latter makes the former disappear, as shown in section \ref{sec:ma}. The 26\,d signal is not consistently strong over the whole time span, although the high cadence data measured after JD2456000 strongly suggests its existence. According to the analyses in \citep{santos14,jenkins14}, this inconsistency of the 26\,d signal is probably caused by stellar activity since the periodogram power at this period is seen in the activity indices such as FWHM. This inconsistency could also be caused by the irregular sampling of the data because signals tend to be more significant in high cadence data. In particular, the time sampling would influence the consistency of long period signals more than that of short period signals because the former need sampling on longer time series to cover their phase. However, this inconsistency of the 26\,d signal is not shown in the MLP-based moving periodogram which does not account for time-varying noise properties. This again demonstrate that the subtraction of a noise component from the data would typically introduce bias in the data analyses. 

To check the consistency of the 290\,d signal in time, we subtract the Keplerian components of the 13.4 and 26\,d signals from the data and show the residual moving periodogram in the right panel of Fig. \ref{fig:BFP-MP}. We see no evidence for strong signals, as indicated by the BF values. On the other hand, the logarithmic BF is relatively high around 18 and 26\,d, but is not significant by being higher than 5. The signals apparent in the residual BFs could arise from the subtraction of signals from all data chunks when these signals do not contribute the same amount of RV variation to different data chunks. This is evident from the non-uniform significance of the 26\,d signal over time shown in the left panel. Considering the potential bias introduced by signal subtraction, the residual moving periodograms can only be used in combination with Bayesian methods which account for all signals together with the noise model parameters simultaneously.

Considering these factors and the analyses shown in section \ref{sec:ma}, we conclude that the 13.4\,d signal corresponds to a planet candidate, and interpret the 18.4\,d signal as an alias of the 13.4\,d signal. Following the conclusion in \cite{jenkins14}, we regard the 26\,d signal as a possible combination of a Keplerian signal and stellar rotation. Moreover, the 13.4 and 26\,d signals are close to the 1:2 period ratio which is found to be common in extra-solar planetary systems \citep{steffen14}. We are suspicious about the existence of the 290\,d signal since we cyannot find it in the residual BFP. We also find that the inclusion of this signal increases the eccentricity of the 26\,d signal and changes the phase of the 13.4\,d signal, indicating that the 290\,d signal is probably a signal connected to the 13.4 and 26\,d signals. Moreover, this signal does not satisfy the logarithmic BF threshold of 5. 

\begin{figure*}
\centering
\includegraphics[scale=0.6]{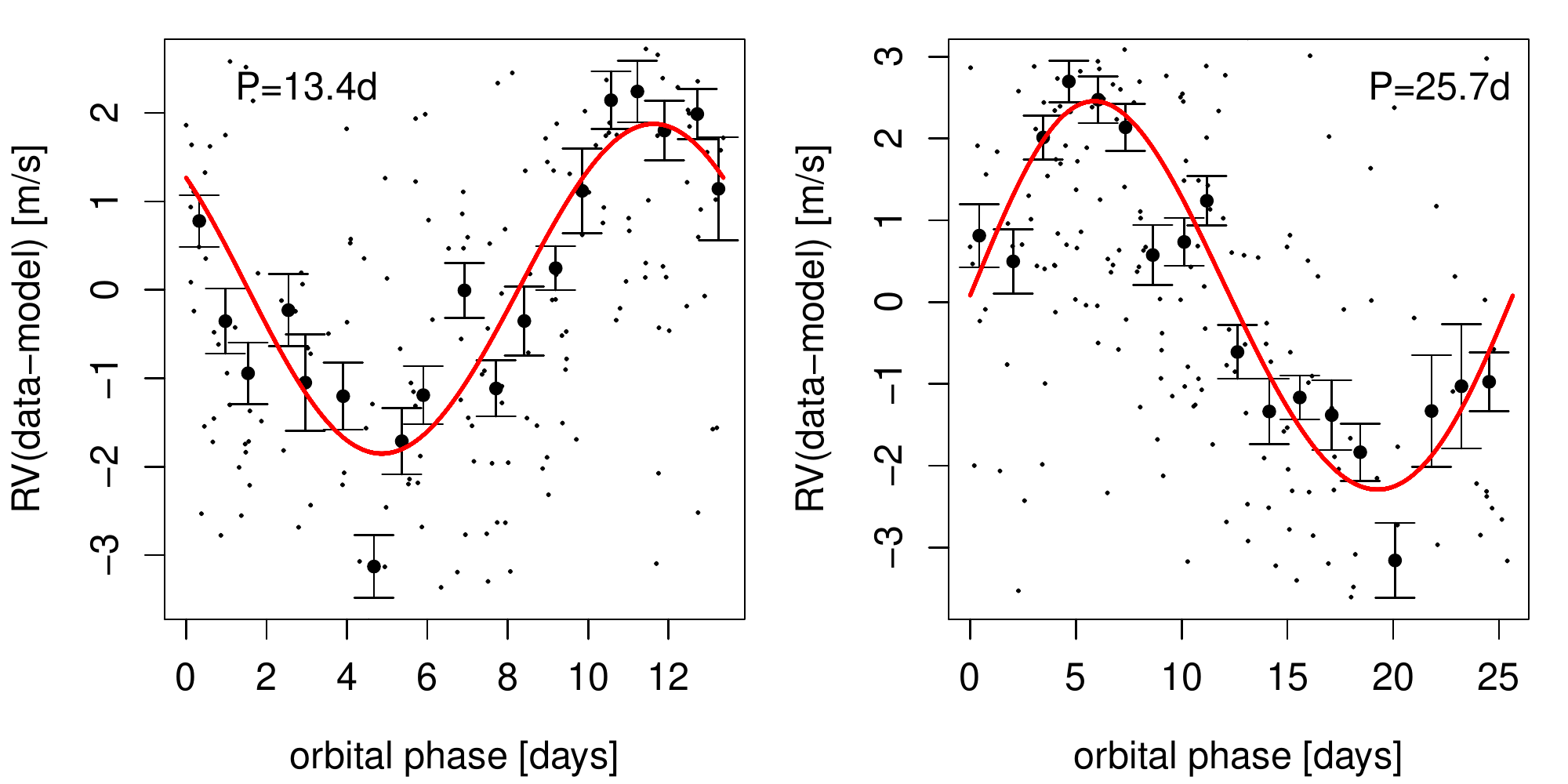}
\caption{As in Fig. \ref{fig:HD177565phase}, but for the HARPS data of HD41248 and the two significant signals in the data.}
\label{fig:HD41248phase}
\end{figure*}

To visualize the fitting of the 13.4 and 26\,d signals quantified by the AM posterior sampling, we show the phase-folded RVs in Fig. \ref{fig:HD41248phase}. We see a good phase coverage, further increasing their credibility as planet candidates. Following \cite{jenkins13}, we adopt a stellar mass of 0.92$^{+0.05}_{-0.05}$\,$M_\odot$ for HD41248, calculate the parameters of these two signals and report them in Table \ref{tab:HD41248sig}. The parameters we estimate for the 26\,d signal are consistent with those found by \cite{jenkins13}. The 13.4\,d planet candidate corresponds to a super-Earth with a minimum mass of 7.08\,$M_\oplus$, and an eccentricity of 0.01, which is lower than the value reported for the 18.4\,d signal by \cite{jenkins13}.  

\begin{table}
  \centering
  \caption{The MAP estimates of the parameters for two signals detected in the TERRA-reduced HARPS data of HD41248. We are dubious about the existence of HD 41248b due to an approximate overlap between its period and rotation period. We estimate the minimum planetary mass and semi-major axis using a stellar mass of 0.92$^{+0.05}_{-0.05}$\,$M_\odot$ \citep{jenkins13}.}
\label{tab:HD41248sig}
  \begin{tabular}  {ccc}
\hline 
Parameters& HD 41248b?& HD 41248c\\\hline
$P$\,(d) &25.654 [25.566, 25.694]&13.365 [13.353, 13.387]\\
$K$\,(m/s) &2.37 [1.30, 2.93]&1.86 [1.03, 2.5]\\
$e$&0.0489 [0.00201. 0.260]&0.0117 [0.00299, 0.257]\\
$\omega$\,(rad) &5.49 [3.26, 9.3]&0.943 [-3.04, 3.04] \\
$M_0$\,(rad) &5.57 [3.22, 9.33]&6.18 [3.24, 9.31]\\
$m\sin{i}$\,($M_\oplus$) &8.36 [5.53, 12.8]&7.08 [3.56, 8.83]\\
$a$\,(au) &0.166 [0.159, 0.173]&0.107 [0.103, 0.112]\\\hline
  \end{tabular}
\end{table}

In summary, the moving periodogram is a useful tool for diagnosing the consistency of signals in time, and in visualizing the change of signals in time. Although the MLP-based moving periodogram is relatively fast to calculate, the BFP-based moving periodogram is more robust in terms of accounting for time-varying noise properties. We expect better performance of the BFP-based moving periodogram due to its ability to account for the trend and correlated noise which vary with observation seasons. The advantage of these new moving periodograms is evident from the analyses of the HARPS data of HD41248. On the other hand, the MCMC-based test of signal consistency is computationally expensive and the divisions of the observational baseline will often be too few for consistency tests. The moving periodograms can provide an alternative and reliable visualization of periodic signals in time series albeit with reliance on careful scaling and visualisation and the need for a long-enough observational baseline.

\section{Analyzing the HARPS data of CoRoT-7 using the Agatha app}\label{sec:agatha}
In this section, we reanalyze the HARPS data of the well-known target, CoRoT-7, and show how to use the Agatha app. The data is obtained from the European Southern Observatory archive, and is processed with the Template-Enhanced Radial velocity Reanalysis Application (TERRA) algorithm \citep{anglada12}. Our data is essentially the same as that used by \cite{tuomi14}. CoRoT-7 is a moderately active G9 star \citep{bruntt10} hosting two planets with orbital periods of 3.7 and 0.85\,d \citep{leger09,queloz09}. There is much controversy over the existence of a third planet with a period of about 9\,d \citep{tuomi14,haywood14,mortier17}. We apply the Agatha app to the reanalysis of the HARPS data of CoRoT-7 as follows. 

\begin{figure*}
    \centering
    \includegraphics[scale=0.6]{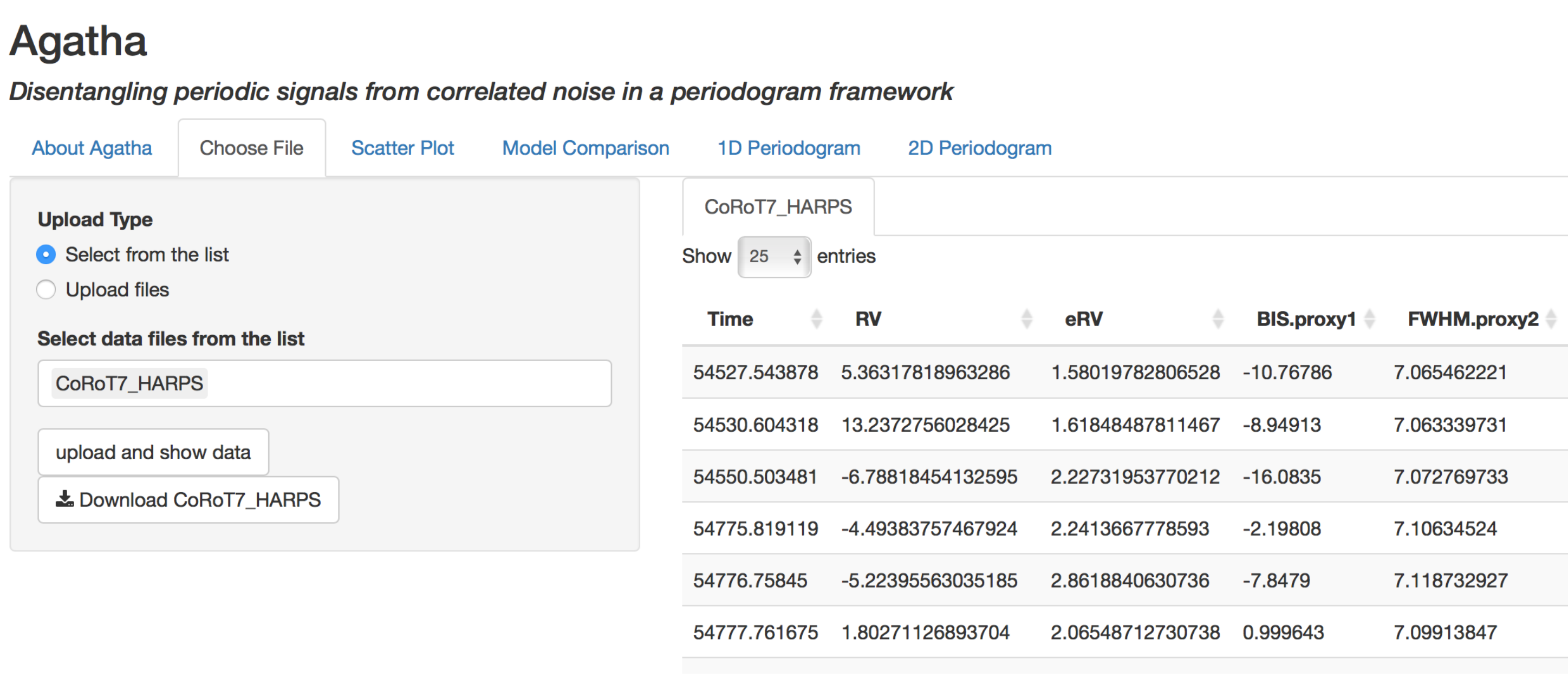}
    \caption{The screen shot of choosing files. }
    \label{fig:choose_file}
    \end{figure*}

First, we choose files from the archived data list or upload our own data. Since we have processed the HARPS data of CoRoT-7 using the TERRA algorithm, we choose it from the list, as seen in Fig. \ref{fig:choose_file}. The data file contains the observation times, RV, RV errors and noise proxies including activity indices, calibration data sets and differential RVs. 

\begin{figure*}
    \centering
    \includegraphics[scale=0.22]{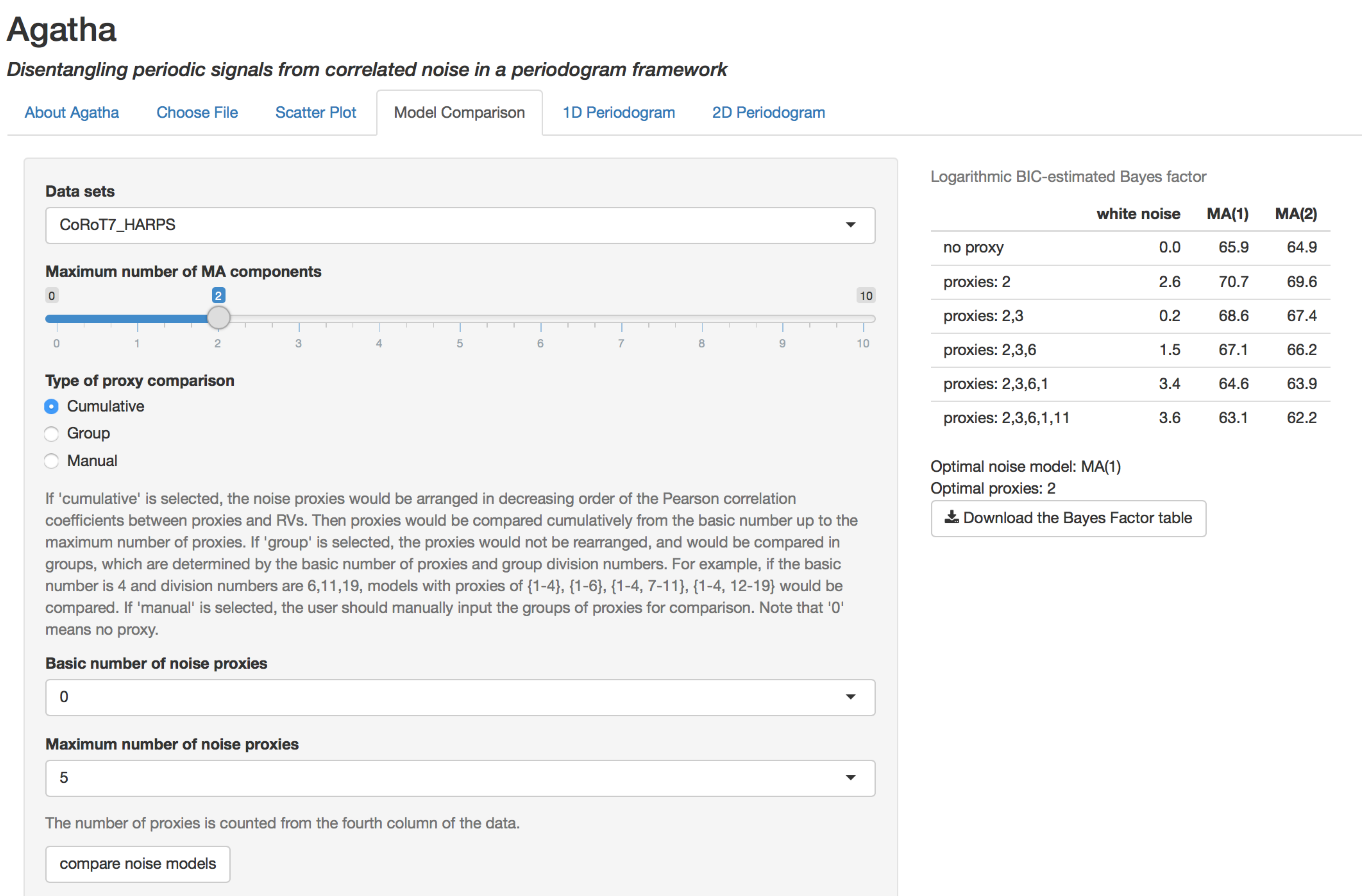}
    \caption{The screen shot of model comparison. }
    \label{fig:model_comparison}
  \end{figure*}

Then, we compare noise models in the ``Model Comparison'' tab. The user can choose the data set for noise model comparison, the maximum number of MA components and the type of proxy comparison. The default proxy comparison is ``Cumulative'', which means that Agatha will first calculate the Pearson correlation coefficients between proxies and RVs. Then Agatha will add proxies one by one in a decreasing order of coefficients to avoid the inclusion of redundant noise proxies which may correlate with the noise proxies included in the model. The basic number of proxies is the number of proxies included in the reference model.  We set the basic number to be zero and set the maximum number of proxies to be 5. We also set the maximum number of MA components to be 2. Then we click the ``compare noise models'' button and get the table of logarithmic BFs as shown in Fig. \ref{fig:model_comparison}. The user can also download the table for publication.

\begin{figure*}
    \centering
    \includegraphics[scale=0.25]{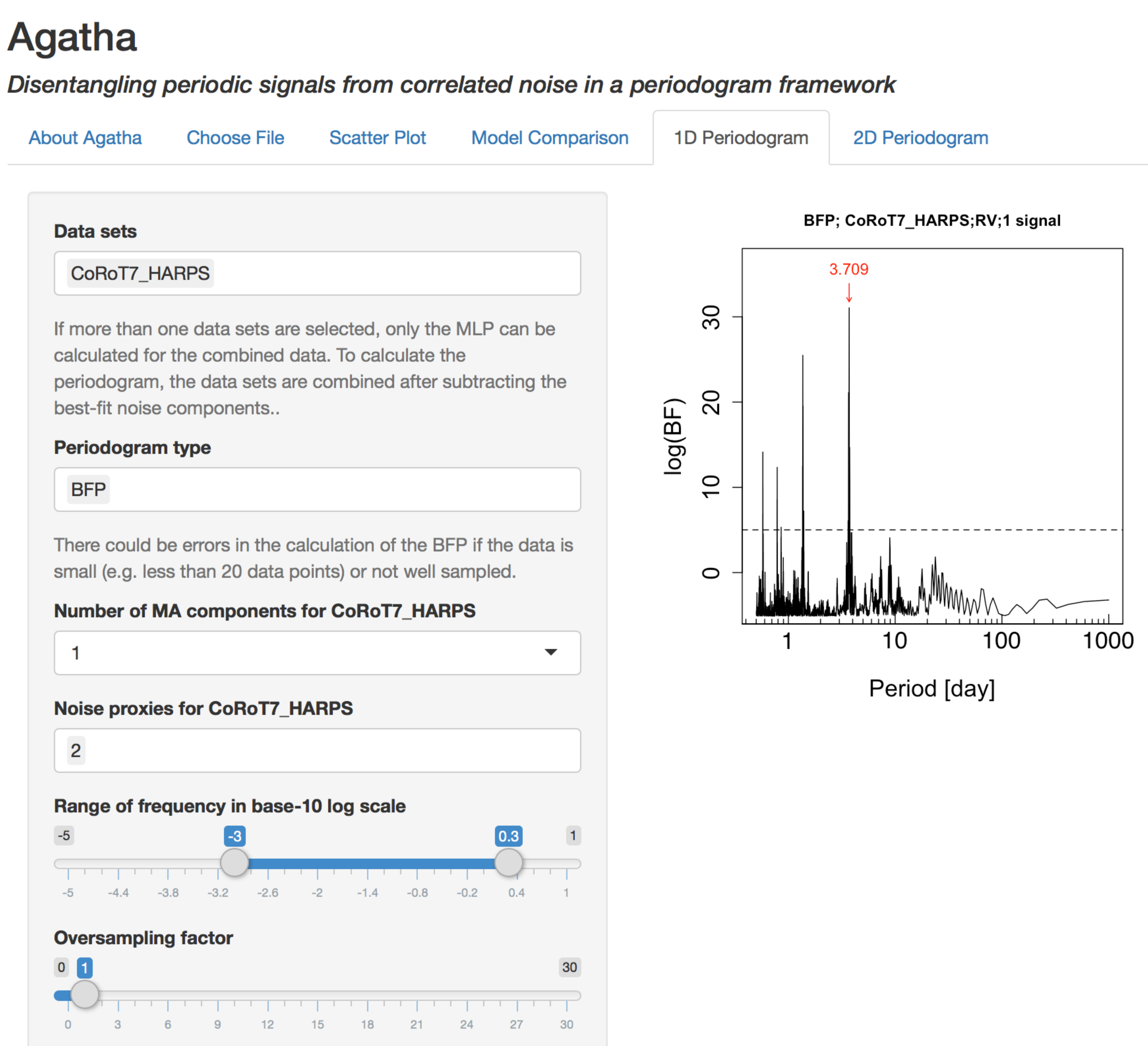}
    \caption{The screen shot of the calculation of BFP. }
    \label{fig:BFP_1sig}
  \end{figure*}

\begin{figure*}
    \centering
    \includegraphics[scale=0.6]{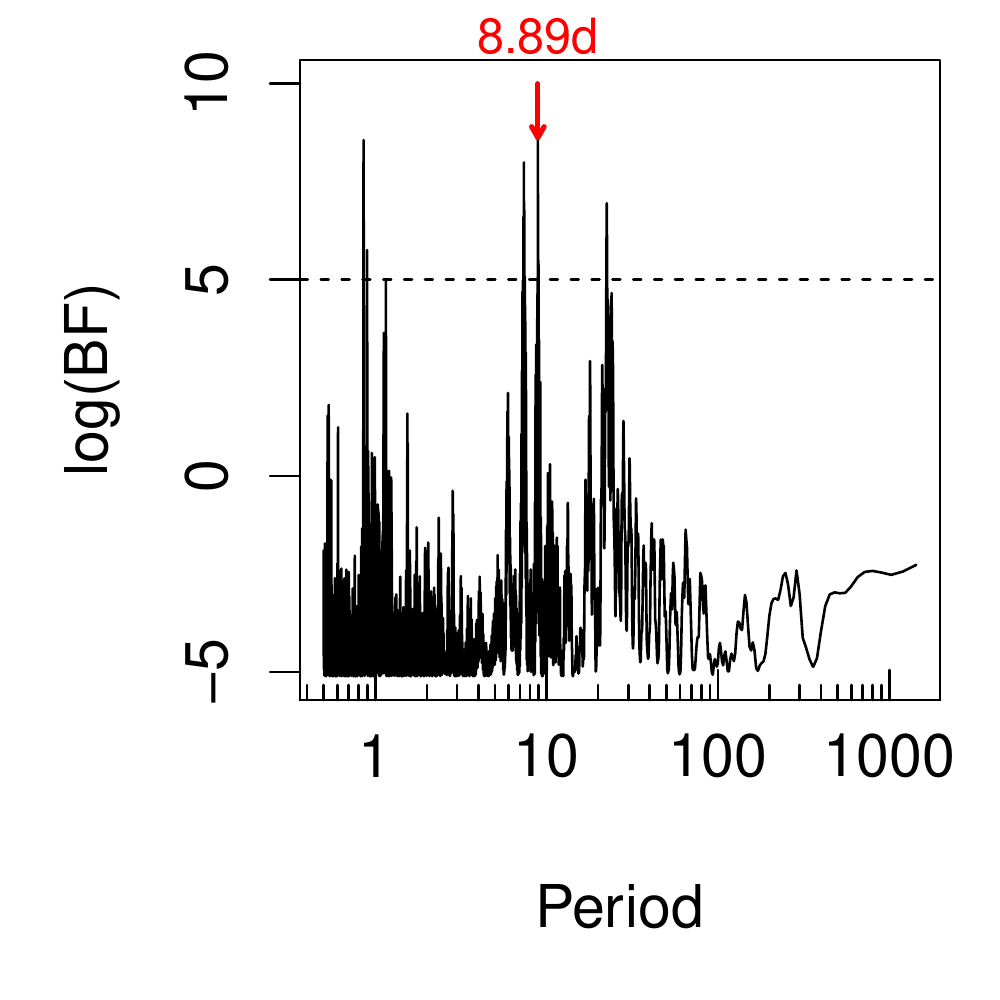}
    \includegraphics[scale=0.6]{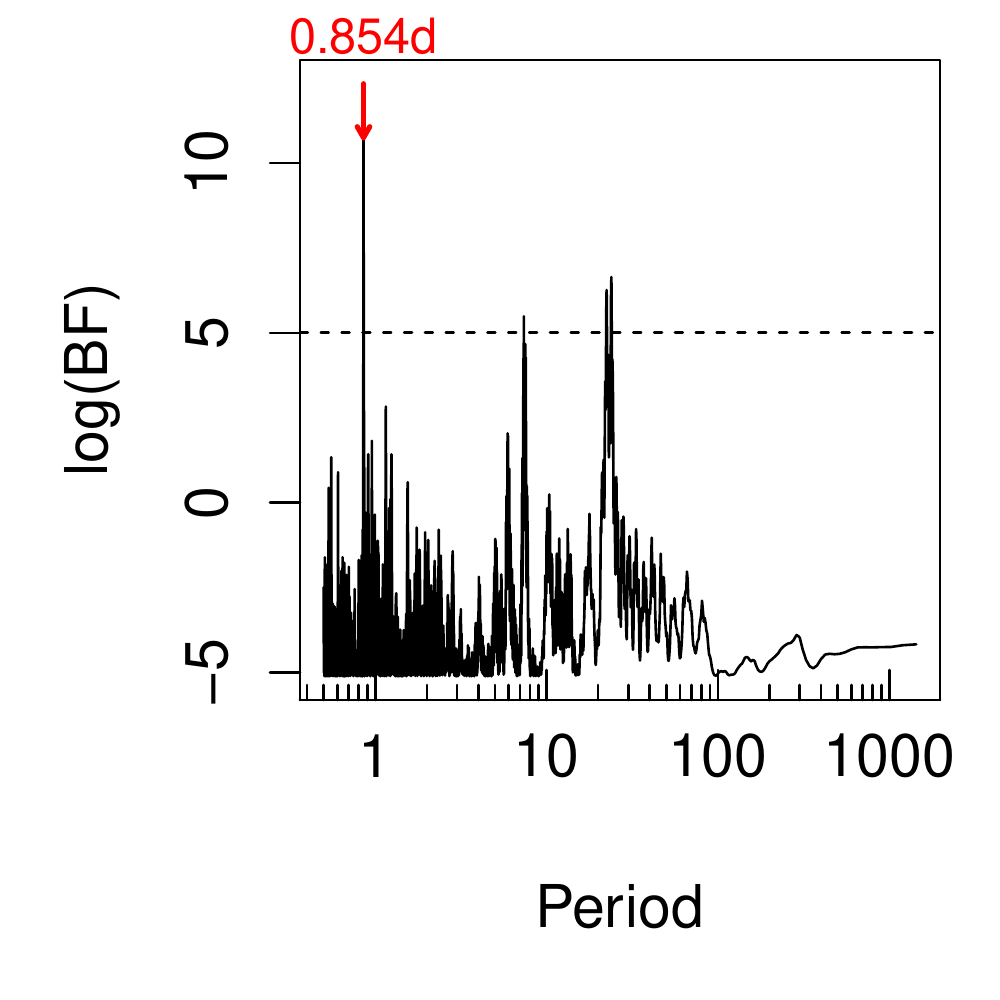}
    \caption{The BFPs of the data subtracted by the Keplerian components of the 1-planet (left) and 2-planet (right) models quantified using MCMC samplings. }
    \label{fig:BFP_residual}
  \end{figure*}

  Based on the logarithmic BF threshold of 5, the optimal noise model is the model which combines MA(1) and a linear function of FWHM. Then we use this noise model in the calculation of periodograms. There are many types of periodograms available, which can be compared if multiple periodograms are chosen. Since the BFP is the most reliable periodogram, we only choose the BFP to visualize signals. We choose the number of MA components and proxies according to the result of model comparison, and click the button of ``plot periodograms'' to calculate the BFP. We show the BFP in Fig. \ref{fig:BFP_1sig}. The logarithmic BF at 3.7\,d is around 30, much larger than the threshold of 5. Hence we quantify this signal using MCMC-based posterior samplings. We calculate the BFP for the data subtracted by the Keplerian component of the 1-planet model and show it in the left panel of Fig. \ref{fig:BFP_residual}. We see a significant signal around 8.9\,d, which is confirmed as a planetary candidate by \cite{tuomi14}. After subtracting this signal from the data, we calculate the residual BFP which is shown in the right panel of Fig. \ref{fig:BFP_residual}. The signal at 0.85\,d is strong in this BFP, which corresponds to the planet detected using the transit method. 

  Therefore the combination of Agatha and MCMC methods can identify three signals at periods of about 0.85, 3.7 and 8.9\,d, which have been confirmed by \cite{tuomi14} using the MA(1) model combined with Bayesian methods. But the 8.9\,d signal is not confirmed by \cite{haywood14} probably due to their usage of Gaussian process, which has been found to interpret signals as noise and thus lead to false negatives \citep{feng16}. Such false negatives could be avoided by using simpler red noise models such as the MA(1) model.  A Bayesian analysis of this data set shows that the inclusion of the 8.9\,d signal increases the BIC-estimated BF by a factor of 25 with respect to the 2-planet model. This provides weak evidence for its existence, although \cite{kass95} consider such an improvement as strong evidence. Moreover, this signal does not overlap with the rotation period which is about 23\,d \citep{queloz09}. But considering that this signal does not pass the logarithmic BF threshold of 5, we don't interpret it as a planet candidate. This signal is not confirmed by \cite{mortier17} either based on the stacked BGLS.

\begin{figure*}
    \centering
    \includegraphics[scale=0.20]{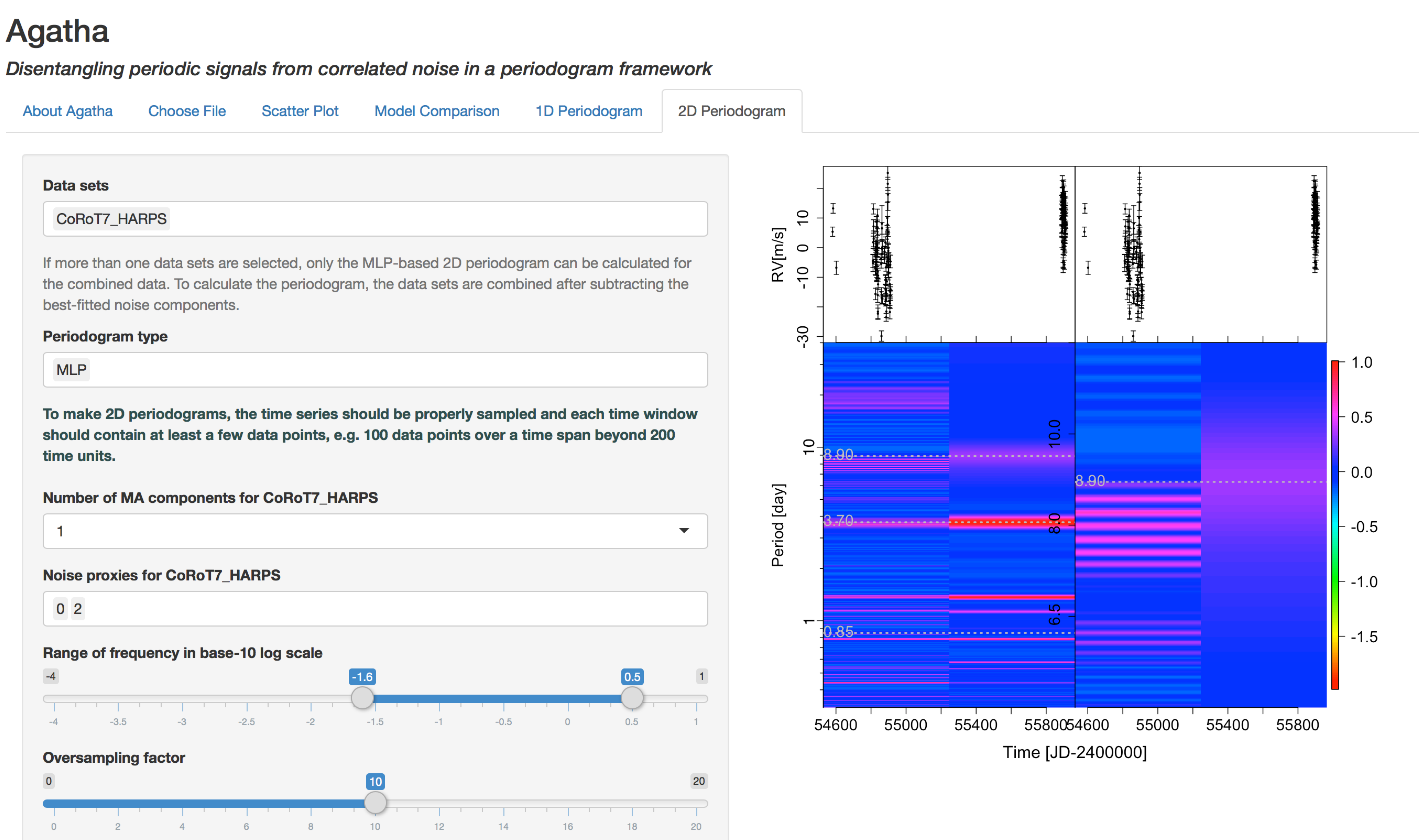}
    \caption{The screen shot of the calculation of moving periodogram. }
    \label{fig:MP_MLP}
  \end{figure*}

  We move on to check the consistency of signals in time using the moving periodogram. We choose the periodogram type, the noise model, the time window, the number of steps and other parameters in the ``2D periodogram'' tab. We calculate the MLP-based moving periodogram in a time window of 300\,d covering the data in 2 steps to avoid the three-year gap between the 2009 and 2012 RV campaigns. The periodogram is shown in Fig. \ref{fig:MP_MLP}. As we can see, the signals at periods of 3.7\,d and 0.85\,d are significant while the 8.9\,d signal is visible over the whole time span. The location of the periodogram power deviates from 0.85\,d probably due to the assumption of circular orbits in the calculation of moving periodogram. Nevertheless, the consistency of signals cannot be fully explored using the moving periodogram because the data size is small and/or the data is not well sampled and so we are not able to be definitive about the consistency of 8.9\,d signal in time.

  In order to make reliable periodograms, the window size and the period range of the moving periodogram should be adjusted according to the property of a given data set. As a rule of thumb, each bin with more than 10 points spread across the bin can be expected to provide constraints on periods less than the bin size. For example, for 100 RVs uniformly sampled over a time span of 1000 day, we recommend a window size of at least 100 day to include at least 10 points in each window for periodogram calculation. If a 100 day window is adopted, the moving periodogram is only able to check the consistency of signals with periods less than 100 day. For RVs sampled with significant gaps, the number of steps could be chosen to avoid the gaps to a large extent. For example, we choose two steps to calculate the MLP for the 2009 and 2012 RV campaigns separately (see Fig. \ref{fig:MP_MLP}). The user should vary the time window and steps to optimize the moving periodogram. 

In summary, we confirm three signals at periods of 0.85, 3.7 and 8.9\,d in the HARPS data of CoRoT-7. Based on our analysis of the current HARPS data, nothing definitive can be said regarding the nature of the 8.9\,d signal. More and well-sampled data is required.

\section{Discussion and conclusions}\label{sec:conclusion}
Complementary to Bayesian methods, Agatha is developed to disentangle periodic signals from correlated noise. Agatha can select the best noise model and assess the significance of periodic signals based on the BIC-estimated BF. Since it optimizes the correlated noise and trend component for each frequency/period, it presents clearer signals than traditional LS periodograms. Moreover, the MLP-based and BFP-based moving periodograms can be used to diagnose the consistency of signals in time, and to reject false positives. 

We test the efficiency of Agatha and the consistency between Agatha and Bayesian methods for an RV challenge data set and the TERRA-reduced HARPS data of HD41248, HD177565 and CoRoT-7. Agatha typically identifies signals consistent with those found by Bayesian methods, although it may miss very weak signals due to its assumption of a single signal. Our analysis of RV challenge system 2 is able to recover signals with signal-to-noise ratio down to $K/N=7.5$, which is probably a limit of reliable detection of planets using the radial velocity method \citep{dumusque16b}. By quantifying the signals identified by Agatha using the AM algorithm for two interesting nearby stars, we find two new planet candidates with 15.1 and 7.08\,$M_\oplus$ orbiting HD177565 and HD41248 with periods of 44.5 and 13.4\,d, respectively. The analysis of the HARPS data of HD41248 shows that the previously claimed 26\,d signal is probably caused by a combination of planetary perturbation and stellar rotation while the 18.4\,d signal is not significant and is an alias of the 13.4\,d signal. We also find clues for other short period signals in the HARPS data of HD177565. We confirm the previous detection of two planets with orbital periods of 0.85 and 3.7\,d in the CoRoT-7 system and provide weak evidence for a signal at a period of 8.9\,d.

Compared with previous red noise periodograms such as RedFit \citep{schulz02} and the compressed sensing periodogram \citep{hara16}, the BFP accounts for signals and noise simultaneously and thus avoids the bias introduced by residual analysis. Instead of assuming white noise as \cite{angus16} did, Agatha accounts for time-correlated noise and is thus more appropriate for analyzing the K2 data to identify stellar rotation periods and to recover acoustic oscillations in giant stars for asteroseismology studies. Moreover, the BFP is the first periodogram that can compare noise models as well as assess signal significance using the BF. To avoid false negatives/positives, the BFP selects the Goldilocks noise model using the BIC-estimated BF. For example, the number of eigen light curves used by \cite{angus16} can be optimized by applying the BFP-based model comparison. Otherwise, the eigen light curves, like differential RVs, might introduce additional noise into model fitting (F17a). Agatha enables wavelength-dependent noise to be assessed which appears to be an important factor in reliable signal detection.

Nevertheless, the application of Agatha is limited by the assumption of single sinusoidal signal. Thus it is not aiming at identify and quantifing signals independently. In addition, we only account for a linear trend in the model while long period signals are better modeled as a second-order polynomial, although such a linear trend sometimes can improve the fitting significantly. We have used linear functions to model the correlation between RVs and noise proxies, which may not be optimal for the nonlinear dependence of RVs on activities as mentioned by \cite{haywood16} and \cite{feng16}. But more sophisticated models may cause the overfitting problem that is mentioned in \cite{feng16}. Hence there is no perfect way to model activity-induced RV variation. 

To reduce the dimensionality of the model comparison, Agatha compares noise models and signals separately. According to our analyses of HARPS data sets (e.g. F17a), the inclusion of periodic functions in the model typically do not change the optimal noise model. Higher order moving average model is only necessary if RVs are measured with high cadence (e.g. \citealt{tuomi12}),  which usually does not contribute to the RV variation induced by planets. Hence the noise and signal selection can be performed separately, although the users can compare the highest BF of BFPs calculated with different noise models to select the optimal combination of signal and noise components. 

Considering that Agatha only estimates the significance of a single sinusoidal signal, we suggest to use it in combination with fully Bayesian methods to find periodic signals. Future developments are necessary to generalize Agatha to identify multiple signals with various periodic functions and red-noise models. Agatha is flexible enough to be adapted for time series analyses in various fields such as paleontology \citep{feng13, melott13, bailer-jones13} and paleoclimatology \citep{wunsch04,lisiecki10,feng15a} and particularly in astronomical fields such as quasar variability \citep{graham15,vaughan16}, stellar activity \citep{reinhold13}, classification of variable stars \citep{richards11}, variability of AGN \citep{hovatta07}, solar cycles \citep{chowdhury13} and other subjects with periodic signals. We list the following examples specifically.  
\begin{itemize}
\item The LS periodogram has been used to identify various periods in the biodiversity variation  (e.g. \citealt{melott12}) while Bayesian methods show no evidence for periodicity \citep{bailer-jones09,feng13}. Instead of subtracting a global trend from the biodiversity time series \citep{melott12}, the BFP accounts for a floating trend and thus avoids potential biases introduced by trend-subtraction.
\item White-noise periodograms \citep{graham15} and Bayesian methods \citep{andrae13} have been used to discover periodicity in quasar light curves. To avoid the over-simplified noise model adopted by traditional periodograms and the computationally expensive posterior samplings used by Bayesian methods, the BFP and MLP can be efficiently calculated for a given quasar light curve to disentangle signals from red noise. In addition, the BFP and MLP-based moving periodograms can be used to visualize the change of quasar periodicity in time. 
\item The LS and GLS have been used to extract periodic features from light curves for the classification of variable stars \citep{richards11,graham13}. To improve the classification of variable stars, the BFP/MLP can be calculated for a given light curve in order to account for the red noise, which is found to be common in various light curves \citep{pont06,aigrain15}. Without consideration of a correlated-noise model, the time-correlated noise would probably lead to false positives and neglecting real signals in the light curves, potentially leading to false classifications of variable stars. 
\item The so-called ``multiband LS periodogram'' has been developed to extract periodic signals from multiband light curves measured by planned multicolor surveys such as LSST \citep{vanderplas15}. However, such a periodogram does not account for wavelength-correlated noise and thus would probably lead to false positives/negatives. Such correlated noise could be modeled by including noise proxies similar to differential RVs in the model.

The Agatha app is made within the Shiny web application framework developed by RStudio (\url{https://shiny.rstudio.com}). It is aimed at visualizing signals reliably in a framework of periodograms rather than finding periodic signals independently. It should be used in combination with MCMC-based posterior samplings to select and quantify multiple signals. Compared with other periodogram-related softwares, the Agatha app is highly interactive and easy to use, without requiring programming skills. It provides the BFP and MLP together with other periodograms for various applications. On the other hand, the widely used exoplanet software, Systemic \citep{meschiari09}, only provides simple periodograms such as GLS and LS, which could be unreliable for planet detections. Period04 \citep{lenz04}, another periodogram software, only use white-noise periodograms, although it can provide multiple-frequency fits. PlanetPack \citep{baluev13}, a C++ software, requires knowledge of C++ for usage and is computationally expensive, though it can deal with time-correlated noise and multiple frequencies. Thus Agatha is a good choice for signal visualization, and would be more versatile if used in combination with Bayesian methods implemented by posterior samplings. Users can use the archived RV data sets or upload their own data to explore signals with Agatha.
\end{itemize}

\section*{Acknowledgements}
FF, MT and HJ are supported by the Leverhulme Trust (RPG-2014-281) and the Science and Technology Facilities Council (ST/M001008/1). We used the ESO Science Archive Facility to collect radial velocity data sets. We would like to thank the anonymous referee for the valuable comments which have enabled significant improvements of the manuscript.

\begin{appendix}
  \section{HARPS data of HD177565 and HD41248}\label{sec:HD177565}
\begin{table*}
  \centering 
  \caption{The HARPS data of HD177565. The online version is available at \url{https://github.com/phillippro/agatha/tree/master/data}. The columns of 3AP2-1 and 3AP3-2 are differential RVs derived from 3 summations of orders. c3AP2-1 is the calibration data.}
  \label{tab:HD177565}
  \footnotesize{
\begin{tabular}{c*{8}{c}}
  \hline
JD-2400000&RV [m/s] &RV error [m/s]& BIS&FWHM&S-index&c3AP2-1 [m/s]&3AP2-1 [m/s]&3AP3-2 [m/s]\\\hline 
52937.514 & 1.22 & 0.43 & -35.73 & 6.8160 & 0.1695 & -3.31 & -3.31 & -1.69 \\ 
  52937.536 & 0.05 & 0.43 & -34.41 & 6.8142 & 0.1683 & -0.25 & -1.45 & -1.27 \\ 
  52939.505 & -0.98 & 0.41 & -37.09 & 6.8150 & 0.1697 & -0.87 & -0.32 & -2.30 \\ 
  52947.501 & 0.10 & 0.40 & -37.99 & 6.8173 & 0.1715 & -2.57 & -3.78 & -3.51 \\ 
  53146.709 & 1.68 & 0.50 & -39.28 & 6.8215 & 0.1705 & -0.13 & 0.59 & -0.05 \\ 
  53146.869 & 0.86 & 0.42 & -38.31 & 6.8175 & 0.1690 & -0.59 & 0.34 & 0.68 \\ 
  53149.911 & 4.07 & 0.50 & -34.40 & 6.8120 & 0.1685 & -0.97 & -1.34 & -0.48 \\ 
  53150.801 & 3.34 & 0.50 & -35.30 & 6.8184 & 0.1761 & -0.19 & 0.66 & -1.08 \\ 
  53154.806 & -2.91 & 0.46 & -36.77 & 6.8131 & 0.1735 & 0.05 & 1.06 & 3.50 \\ 
  53201.701 & -3.40 & 1.64 & -35.00 & 6.8071 & 0.1539 & 0.29 & 6.53 & 7.56 \\ 
  53201.705 & -0.89 & 2.06 & -29.77 & 6.8115 & 0.1526 & 3.63 & 3.63 & 5.97 \\ 
  53201.710 & -2.62 & 2.32 & -31.83 & 6.8018 & 0.1534 & 0.24 & -2.62 & 8.56 \\ 
  53202.682 & 0.93 & 0.36 & -36.53 & 6.8128 & 0.1709 & -0.48 & -0.39 & 1.13 \\ 
  53202.688 & 1.21 & 0.37 & -37.39 & 6.8152 & 0.1711 & -0.53 & -1.19 & 0.25 \\ 
  53203.699 & -0.30 & 0.27 & -36.28 & 6.8213 & 0.1706 & -0.76 & -4.20 & 1.68 \\ 
  53203.704 & 1.85 & 0.28 & -36.58 & 6.8192 & 0.1710 & -1.07 & -4.05 & 0.49 \\ 
  53204.660 & -4.61 & 0.35 & -36.35 & 6.8148 & 0.1706 & 0.72 & 3.20 & 2.44 \\ 
  53204.663 & -5.72 & 0.34 & -36.75 & 6.8146 & 0.1695 & 3.57 & 3.57 & 3.37 \\ 
  53204.667 & -5.13 & 0.36 & -34.60 & 6.8134 & 0.1705 & 0.21 & 2.54 & 2.58 \\ 
  53205.577 & 0.00 & 0.50 & -37.32 & 6.8159 & 0.1691 & 0.27 & -1.68 & -1.41 \\ 
  53205.579 & 1.73 & 0.54 & -34.97 & 6.8140 & 0.1692 & 0.40 & 0.40 & 0.99 \\ 
  53205.581 & 0.19 & 0.50 & -34.96 & 6.8184 & 0.1700 & -0.79 & -0.79 & -1.14 \\ 
  53205.582 & -0.24 & 0.56 & -35.76 & 6.8140 & 0.1668 & 1.48 & 1.48 & -2.10 \\ 
  53205.584 & 0.50 & 0.55 & -37.78 & 6.8145 & 0.1674 & 0.87 & 0.40 & -0.39 \\ 
  53206.682 & -1.80 & 0.32 & -36.29 & 6.8105 & 0.1698 & 0.17 & 0.70 & 0.89 \\ 
  53206.686 & -0.69 & 0.32 & -35.50 & 6.8114 & 0.1699 & -0.99 & -0.99 & 2.97 \\ 
  53206.689 & -1.24 & 0.32 & -36.17 & 6.8135 & 0.1705 & -0.29 & -0.52 & 0.43 \\ 
  53217.651 & -7.59 & 0.42 & -37.65 & 6.8090 & 0.1684 & 0.15 & 4.04 & 1.14 \\ 
  53217.655 & -7.40 & 0.42 & -36.01 & 6.8078 & 0.1679 & 0.80 & 0.80 & 0.84 \\ 
  53217.659 & -7.11 & 0.44 & -35.99 & 6.8096 & 0.1685 & -0.05 & 0.82 & 2.42 \\ 
  53218.686 & -0.24 & 0.40 & -36.36 & 6.8140 & 0.1692 & -0.86 & -1.45 & -0.44 \\ 
  53218.690 & 0.71 & 0.39 & -35.94 & 6.8157 & 0.1678 & -1.20 & -1.20 & 0.15 \\ 
  53218.694 & 0.11 & 0.41 & -37.54 & 6.8143 & 0.1679 & 0.54 & 0.26 & -0.49 \\ 
  53230.669 & 2.75 & 0.38 & -36.15 & 6.8193 & 0.1703 & 0.04 & -1.39 & -1.62 \\ 
  53230.673 & 2.08 & 0.36 & -36.02 & 6.8188 & 0.1694 & -1.33 & -1.33 & 0.92 \\ 
  53230.677 & 0.71 & 0.38 & -35.96 & 6.8190 & 0.1711 & 1.22 & -0.66 & 0.99 \\ 
  53232.626 & 1.40 & 0.62 & -36.34 & 6.8173 & 0.1689 & 1.18 & -0.01 & -0.03 \\ 
  53232.630 & 3.14 & 0.67 & -34.29 & 6.8149 & 0.1651 & 2.19 & 2.19 & -3.03 \\ 
  53232.634 & 2.21 & 0.62 & -37.20 & 6.8202 & 0.1685 & -1.25 & -0.83 & -0.32 \\ 
  53263.578 & -3.23 & 0.28 & -37.47 & 6.8209 & 0.1681 & -1.74 & -2.07 & 1.16 \\ 
  53263.583 & -1.14 & 0.29 & -37.65 & 6.8226 & 0.1682 & -2.04 & -2.84 & 0.89 \\ 
  53264.553 & -4.67 & 0.29 & -36.51 & 6.8191 & 0.1681 & 0.51 & 0.93 & 3.79 \\ 
  53264.559 & -5.37 & 0.29 & -37.03 & 6.8175 & 0.1684 & -0.85 & 1.84 & 4.49 \\ 
  53269.565 & -0.48 & 0.53 & -38.85 & 6.8169 & 0.1676 & -0.02 & 2.38 & -1.10 \\ 
  53269.570 & 1.01 & 0.68 & -35.30 & 6.8179 & 0.1653 & 0.93 & 2.80 & -0.07 \\ 
  53551.710 & 2.18 & 0.30 & -33.61 & 6.8159 & 0.1808 & -0.24 & 0.29 & -1.09 \\ 
  53817.909 & 4.71 & 0.33 & -30.88 & 6.8233 & 0.1859 & 0.97 & -0.12 & -2.71 \\ 
  53817.914 & 4.58 & 0.31 & -30.20 & 6.8222 & 0.1856 & 0.79 & 0.79 & -2.60 \\ 
  54257.771 & 2.86 & 0.73 & -29.54 & 6.8394 & 0.1976 & 0.76 & 0.43 & 4.12 \\ 
  54257.772 & 4.72 & 0.69 & -29.04 & 6.8400 & 0.1930 & -0.56 & -0.56 & -0.17 \\ 
  54257.774 & 4.28 & 0.68 & -29.43 & 6.8409 & 0.1953 & 3.33 & 3.33 & -1.79 \\ 
  54257.776 & 4.22 & 0.65 & -30.73 & 6.8412 & 0.1925 & 0.59 & 0.59 & -1.52 \\ 
  54257.777 & 3.84 & 0.64 & -28.73 & 6.8405 & 0.1943 & -0.60 & -0.27 & -0.91 \\ 
  54258.854 & 4.13 & 0.68 & -28.24 & 6.8444 & 0.1922 & 1.26 & 1.44 & -3.34 \\ 
  54258.856 & 3.22 & 0.75 & -28.70 & 6.8475 & 0.1932 & 0.97 & 0.97 & -0.77 \\ 
  54258.858 & 3.77 & 0.73 & -30.18 & 6.8435 & 0.1945 & 0.09 & 0.09 & -1.63 \\ 
  54258.860 & 4.90 & 0.72 & -31.25 & 6.8448 & 0.1932 & 3.05 & 3.05 & -0.55 \\ 
  54258.863 & 3.00 & 0.68 & -30.68 & 6.8420 & 0.1929 & -0.18 & 1.58 & 0.99 \\ 
  54259.913 & 4.37 & 1.31 & -31.23 & 6.8498 & 0.1867 & 2.17 & -1.55 & 1.79 \\ 
  54259.916 & 6.43 & 1.68 & -24.13 & 6.8417 & 0.1839 & 2.17 & 13.06 & -7.10 \\ 
  54259.918 & 4.30 & 2.17 & -28.72 & 6.8426 & 0.1853 & -0.29 & -0.29 & 0.80 \\ 
  54259.921 & -2.57 & 2.73 & -34.71 & 6.8590 & 0.1637 & -0.29 & 11.41 & 11.11 \\ 
  54259.924 & 2.63 & 1.86 & -28.04 & 6.8483 & 0.1761 & 4.30 & 1.68 & -0.13 \\ 
  54292.689 & 0.43 & 0.33 & -28.57 & 6.8325 & 0.1933 & 1.07 & 1.51 & -1.01 \\ 
  54292.695 & 2.55 & 0.33 & -28.36 & 6.8329 & 0.1927 & -0.99 & 0.81 & -2.48 \\ 
  54617.853 & 5.46 & 0.72 & -35.71 & 6.8369 & 0.1843 & 0.88 & -0.02 & -3.61 \\ 
  54617.858 & 5.15 & 0.82 & -30.05 & 6.8382 & 0.1839 & 1.31 & 1.92 & -4.95 \\ 
  54621.899 & 2.27 & 0.30 & -33.66 & 6.8294 & 0.1820 & -0.06 & -0.52 & -0.43 \\ \hline
\end{tabular}
}
\end{table*}

\begin{table*}
  \centering 
  \caption{Similar to Table \ref{tab:HD177565} but for HD41248. The full data set is put online. } 
  \label{tab:HD41248}
  \footnotesize{
\begin{tabular}{c*{8}{c}}
  \hline
JD-2400000&RV [m/s] &RV error [m/s]& BIS&FWHM&S-index&c3AP2-1 [m/s]&3AP2-1 [m/s]&3AP3-2 [m/s]\\\hline 
52943.85 & -1.89 & 2.13 & -35.93 & 6.72 & 0.17 & -1.29 & -1.91 & 11.42 \\ 
  52989.71 & -7.34 & 3.22 & -27.40 & 6.72 & 0.16 & -2.43 & -11.29 & 15.96 \\ 
  52998.69 & -0.22 & 3.99 & -33.53 & 6.70 & 0.15 & -0.62 & 15.95 & -16.83 \\ 
  53007.68 & -2.34 & 2.21 & -28.61 & 6.72 & 0.17 & 3.50 & 3.50 & 2.23 \\ 
  53787.61 & -4.41 & 2.63 & -31.31 & 6.72 & 0.16 & 0.89 & -12.67 & 0.53 \\ 
  54055.84 & -5.49 & 2.02 & -23.95 & 6.71 & 0.17 & 1.26 & 2.67 & 4.35 \\ 
  54789.72 & -4.79 & 0.85 & -27.43 & 6.72 & 0.18 & 0.27 & -1.53 & 2.25 \\ 
  54790.69 & -7.22 & 0.93 & -30.83 & 6.72 & 0.17 & 0.67 & 0.67 & 0.42 \\ 
  54791.71 & -4.86 & 0.86 & -29.54 & 6.72 & 0.18 & -2.48 & -0.98 & 1.34 \\ 
  54792.70 & -5.02 & 0.82 & -28.09 & 6.73 & 0.18 & 0.08 & 0.08 & 0.82 \\ 
  54793.72 & -3.35 & 0.92 & -25.28 & 6.73 & 0.18 & -2.38 & 0.04 & 1.53 \\ 
  54794.69 & 0.04 & 0.91 & -29.59 & 6.73 & 0.18 & 2.71 & 2.71 & -2.06 \\ 
  54795.72 & 0.48 & 0.93 & -29.54 & 6.73 & 0.18 & 0.75 & 0.75 & -0.54 \\ 
  54796.72 & 0.51 & 0.98 & -30.33 & 6.73 & 0.18 & -1.36 & -1.36 & 2.69 \\ 
  54797.71 & 2.19 & 0.93 & -27.77 & 6.73 & 0.18 & -1.72 & 1.31 & -1.64 \\ 
  54798.70 & 3.18 & 0.93 & -25.14 & 6.73 & 0.18 & -0.06 & -4.92 & -3.24 \\
  ...&... &... &... &... &... &... &... &... \\
\hline
\end{tabular}
}
\end{table*}

\end{appendix}

\bibliographystyle{aasjournal}
\bibliography{nm}
\end{document}